\documentclass[%
 reprint,
]{revtex4-2}
\usepackage{graphicx}
\usepackage{dcolumn}
\usepackage{bm}
\usepackage{amsmath}
\usepackage{caption}
\usepackage{subcaption}
\usepackage{color}
\usepackage{siunitx}

\begin{document}

\preprint{AIP/123-QED}

\title[Ultra-large polymer-free suspended graphene films]{Ultra-large polymer-free suspended graphene films}

\author{L.~Kalkhoff\thinspace$^a$$^b$}
 \email{lukas.kalkhoff@uni-due.de}
\author{S.~Matschy\thinspace$^{a}$}
\author{A.S.~Meyer\thinspace$^a$$^b$}
\author{L.~Lasnig\thinspace$^a$$^b$}
\author{N.~Junker\thinspace$^a$$^b$}
\author{M.~Mittendorff\thinspace$^a$}
\author{L.~Breuer\thinspace$^a$$^b$}%
\author{M.~Schleberger\thinspace$^a$$^b$}%
\email{marika.schleberger@uni-due.de}
 \affiliation{University of Duisburg Essen, Physics Department\thinspace$^a$ and CENIDE\thinspace$^b$, Lotharstr. 1, 47057 Duisburg, Germany}

\date{\today}

\begin{abstract}
Due to its extraordinary properties, suspended graphene is a critical element in a wide range of applications. Preparation methods that preserve the unique properties of graphene are therefore in high demand. To date, all protocols for the production of large graphene films have relied on the application of a polymer film to stabilize graphene during the transfer process. However, this inevitably introduces contaminations that have proven to be extremely difficult, if not impossible, to remove entirely. Here we report the polymer-free fabrication of suspended films consisting of three graphene layers spanning circular holes of 150~$\mathrm{\text\textmu m}$ diameter. We find a high fabrication yield, very uniform properties of the freestanding graphene across all holes as well across individual holes. A detailed analysis by confocal Raman and THz spectroscopy reveals that the triple-layer samples exhibit structural and electronic properties similar to those of monolayer graphene. We demonstrate their usability as ion-electron converters in time-of-flight mass spectrometry and related applications. They are two orders of magnitude thinner than previous carbon foils typically used in these types of experiments, while still being robust and exhibiting a sufficiently high electron yield. These results are an important step towards replacing free-standing ultra-thin carbon films or graphene from polymer-based transfers with much better defined and clean graphene. 
\end{abstract}

\maketitle

\section{\label {Introduction} Introduction\protect}

Ultrathin carbon films are key materials for many applications and experiments. Supported films are used for example as coating for battery electrodes \cite{Li.2012} or as sensors for virus detection~\cite{Adeel.2022}. A very important and large field of applications stems from the fact that ion irradiation of solids leads to the emission of electrons~\cite{Oliphant.1930b,Oliphant.1930}.
Thus, supported carbon films are, e.g., used in space instrumentation for incident ion detection~\cite{Ebert.2014,Allegrini.2003,Allegrini.2016,Vira.2020}, while suspended carbon foils are frequently used in coincidence- and time-of-flight (ToF) measurements~\cite{Menendez.1986,McComas.1990,Holenak.2021}, or for stripping purposes, i.e.~the ionization of energetic neutral atoms and charge equilibration of ions~\cite{Moak.1976,Hattass.1999}. 

With the discovery of graphene (Gr), it became for the first time possible to prepare suspended carbon films of high structural quality~\cite{Meyer.2007} that are two orders of magnitude thinner than the thinnest conventional carbon foils. This has led to even more advanced applications such as ultrasensitive pressure sensors and actuators~\cite{Chen.2016,Roson.2020,Carvalho.2022,Romijn.2021,Shin.2023} windows for \textit{in vivo}/\textit{in situ} scanning electron microscopy~\cite{Stoll.2012} as well as X-ray photoelectron spectroscopy~\cite{Leidinger.2021}, and membranes for ultrafiltration~\cite{OHern.2012,Koenig.2012,Surwade.2015,Madau.2017}. 

Suspended graphene is in particular well-suited for studying and exploiting ion-solid interactions, as it exhibits a high structural quality and a well-defined thinness. Further, it has not only been shown to be mechanically robust~\cite{Frank.2007,Lee.2008,Booth.2008} and extremely radiation hard~\cite{Ochedowski.2013,Kozubek.2014,Ernst.2016,Gruber.2016,Kuramitsu.2022} but it also features an outstanding self-healing capability~\cite{Zan.2012} making it superior to any other material. Thus it comes as no surprise, that exfoliated graphene, suspended on transmission electron microscopy (TEM) grids has helped to significantly advance our understanding of ion-solid-interactions~\cite{Gruber.2016, Creutzburg.2021}.

However, the samples used in these studies have been extremely small and the fabrication of suspended graphene targets, that are sufficiently large (i.e. much larger than the areas typical for a TEM grid) and clean, poses a serious challenge. In particular, polymers, which are frequently used in the preparation of large suspended graphene samples (see e.g.~ref.~\cite{Chen.2016}), cannot be completely removed, and even worse, often form a homogeneous film, as recently shown by Tilmann \textit{et al}.~\cite{Tilmann.2023}. This can however not be tolerated in most cases, as e.g. these unavoidable residues would prevent the interaction of ions with graphene. Furthermore, for any time-sensitive experiment, the samples also need to be extremely flat in order not to negatively affect the time-resolution as will be discussed in more detail below. 

In this paper we report how such suspended large-area, high quality graphene targets can be fabricated from commercially available graphene grown by chemical vapour deposition (CVD) without the use of polymers. We present a detailed study of their structural and electronic properties revealed by optical profilo\-metry as well as confocal Raman- and THz spectroscopy. Finally, we show that these targets despite their thinness can be used as efficient ion-to-electron converters and are thus indeed suitable targets for fundamental research and as well for applications in the field of ion-solid-interaction.


\section{\label{SamplePrep}Preparation of large area, contamination free graphene\protect}
To avoid the contamination of our samples we have established a completely polymer-free wet transfer process.
The preparation process starts by applying a procedure that we have developed earlier to increase the mechanical stability of graphene and funtionalize it (for details see~\cite{Madau.2020}). Briefly, a Gr/Cu wafer produced by CVD is placed into an ethylene-vinyl acetate co-polymer foil together with a supporting solid, smooth plate. The Gr/Cu wafer package is then put into a plastic bag that is evacuated and sealed with a commercially available vacuum sealer. The Gr/Cu wafer stays in contact with the co-polymer foil under these vacuum conditions for at least 24~hrs. After the seal is broken, the wafer is cut into pieces of roughly 10~mm~x~10~mm and placed into an ammonium persulfate solution of 20\thinspace\% concentration. After the copper is fully dissolved, the etching solution is diluted by de-ionized (DI) water and the graphene floating on the surface is scooped onto another Gr/Cu piece.

After the stack is fully dry, this process is repeated until the desired number of graphene layers is obtained. For reasons which will become apparent in below, we will refer to these as double (and not bilayer), triple or quadruple layers. In the final step, the stack is etched in the manner previously described, then transferred onto a substrate of 250~$\mathrm{\text\textmu m}$ thick silicon (Si), or steel plate, featuring a periodic matrix of 200 holes, with a diameter of 150~$\mathrm{\text\textmu m}$ each, created by laser ablation. 

\begin{figure}[ht]
       \includegraphics[width=0.5\textwidth]{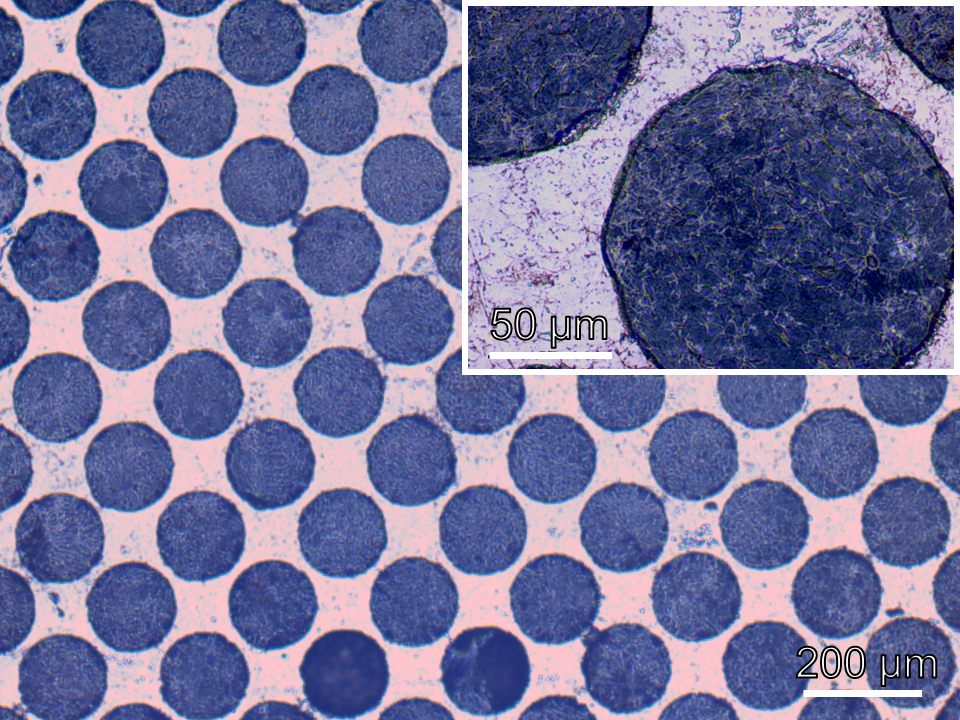} 
       \captionsetup{justification=raggedright,singlelinecheck=false}
       \caption{Optical microscopy image from a a triple layer graphene transferred onto a Si substrate with an array of 150~$\mathrm{\text\textmu m}$ holes with 5x magnification and 50x magnification in the inset on the top right.}
   \label{Opt}
\end{figure}

In principle, targets with double layers of graphene can be manufactured in this way, but then the coverage of holes is insufficient. This might be due to the fact that CVD graphene is known to have a reduced mechanical stability compared to exfoliated graphene because of more grain boundaries~\cite{Huang.2011}. These grain boundaries are susceptible to tearing if not reinforced by an additional layer. In the case of double layer graphene, the likelihood of two mechanically weak points aligning may still be significant enough to result in a reduced coverage. 

In contrast, triple layer graphene stacks exhibit sufficient stability, as evidenced by the freestanding graphene layers spanning 150~$\mathrm{\text\textmu m}$ shown in Fig~\ref{Opt}. The overall coverage is typically $>$90\thinspace\% for triple layers, and with additional layers, an even higher coverage can be achieved. In the following, we will focus on triple layers of graphene, as this represents a good compromise between minimum thickness and a sufficiently high coverage.

\section{\label{Results} Properties \protect}
\subsection{Optical Profilometry: Topography}
\begin{figure}[ht]
    \centering
    \includegraphics[width=0.5\textwidth]{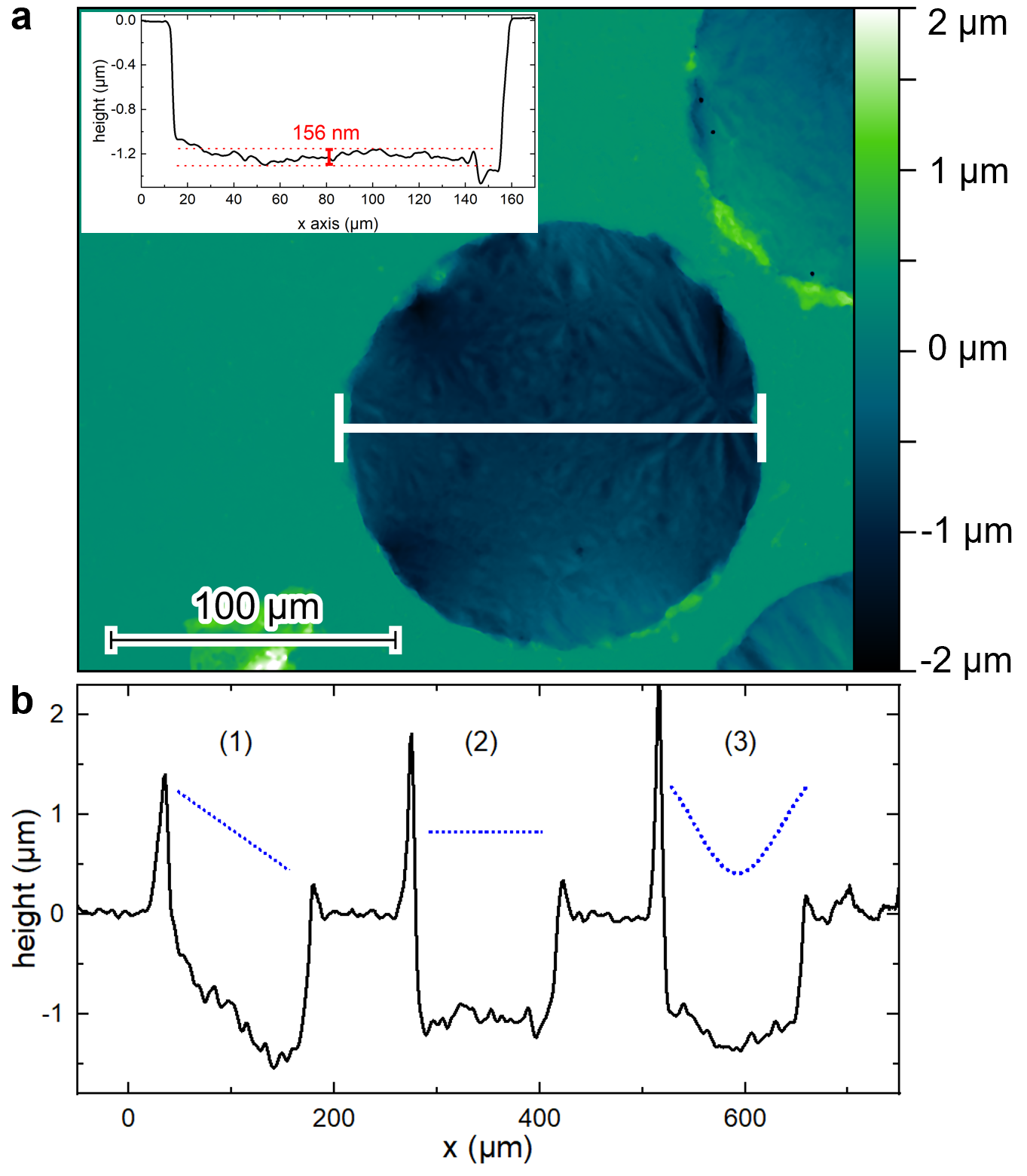}
    \captionsetup{justification=raggedright,singlelinecheck=false}
     \caption{\textbf{a} Optical profilometry of freestanding triple layer graphene spanning a hole with a diameter of 150~$\mathrm{\text\textmu m}$. The inset shows the height distribution of a line scan marked in the image by the white line with a height difference of 156~nm. A line scan spanning over three holes is seen in \textbf{b} with the most common types of graphene topography within the holes which are simplified by the blue dotted lines. Type (1) has a steady slope, type (2) is plane-parallel to the surface, type (3) is hammock-like.}
   \label{Optical}
\end{figure}

To obtain a non-destructive representation of the topography, we utilized a 3D optical profilometer (S Neox 090 Sensofar). A typical false color image can be seen in Fig.~\ref{Optical}~\textbf{a}. In this example, graphene covers the complete hole of 150~$\mathrm{\text\textmu m}$ and is almost plane-parallel to the Si-substrate surface as can be seen from the line scan in the inset of Fig.~\ref{Optical}~\textbf{a}. With the help of such line scans we could identify two more topography types within the holes as depicted by the blue dotted lines in Fig.~\ref{Optical}~\textbf{b}. 

These three types of topography result in distinctive height differences. Type (1) shows a slope with the largest height difference, while type (2) is plane parallel and results in the smallest possible one. Type (3), the hammock-like, results in an intermediate height difference. This is an important quantity for samples used as a target material in any time-sensitive experiment because one needs to know the exact starting point of the emitted particles in order to evaluate their flight time with respect to a given time zero. 

To illustrate this more clearly, let us look at an example. Ultra thin carbon foils are frequently used to generate a START-signal for coincidence- or ToF measurements because ions cause electron emission when passing through such foils~\cite{Baragiola.1998}. If we would use a graphene target like the one seen in Fig.~\ref{Optical}~\textbf{a}, with the maximal height deviation of 156~nm, an ion impact would thus in the worst case take place either at the highest point of the graphene target or at the lowest, respectively. For a pulse of Ne$^+$ ions with 3.5~keV energy, e.g., this would result in a time difference $\Delta t$ for the arrival of the ions and thus the emission of any secondary particles corresponding to: 

\begin{equation*}
\centering
\Delta t = \Delta h \cdot \sqrt{\frac{m_{Ne}}{2E_{kin}}} \simeq 156~\mathrm{nm} \cdot \sqrt{\frac{20~\mathrm{u}}{2\cdot 3.5~\mathrm{keV}}} \simeq 0.85~\mathrm{ps}
\label{eq1}
\end{equation*}

This is well below the time-resolution of any current experiment and thus, graphene targets of type (2) are the best case scenario, as all emitted particles would practically start at the same time after the ion impact. Graphene targets of type (1) exhibit a maximum height difference of over 1~$\mathrm{\text\textmu m}$ which seems large in light of the discussion above. However, it is rather easy to correct for this difference by tilting the graphene surface to become plane-parallel. After plane-subtraction the remaining mean height variation for this type is typically on the order of (100--300)~nm, similar to type (2). In contrast, type (3) targets resemble a "hammock", for which there is no way of correction possible and therefore are not suitable for time critical experiments. However, with only 4\thinspace\% occurrence this type is relatively rarely seen in our samples with a fully covered target, while 42\thinspace\% are of type (2) and 54\thinspace\% of type (1), respectively.  

\subsection{Raman: Structure, Strain and Doping}
High quality films should not just be flat but should also feature similar properties over the whole sample area. We therefore probed the homogeneity of graphene within the holes using Raman spectroscopy. The positions of the two characteristic modes of graphene, namely the $G$ and $2D$ mode allow for a determination of layer number, charge carrier density, and mechanical strain as shown by Lee \textit{et al}~\cite{Lee.2012} and Kim \textit{et al}.~\cite{Kim.2016}. 

\begin{figure}
     \centering  
        \includegraphics[width=0.5\textwidth]{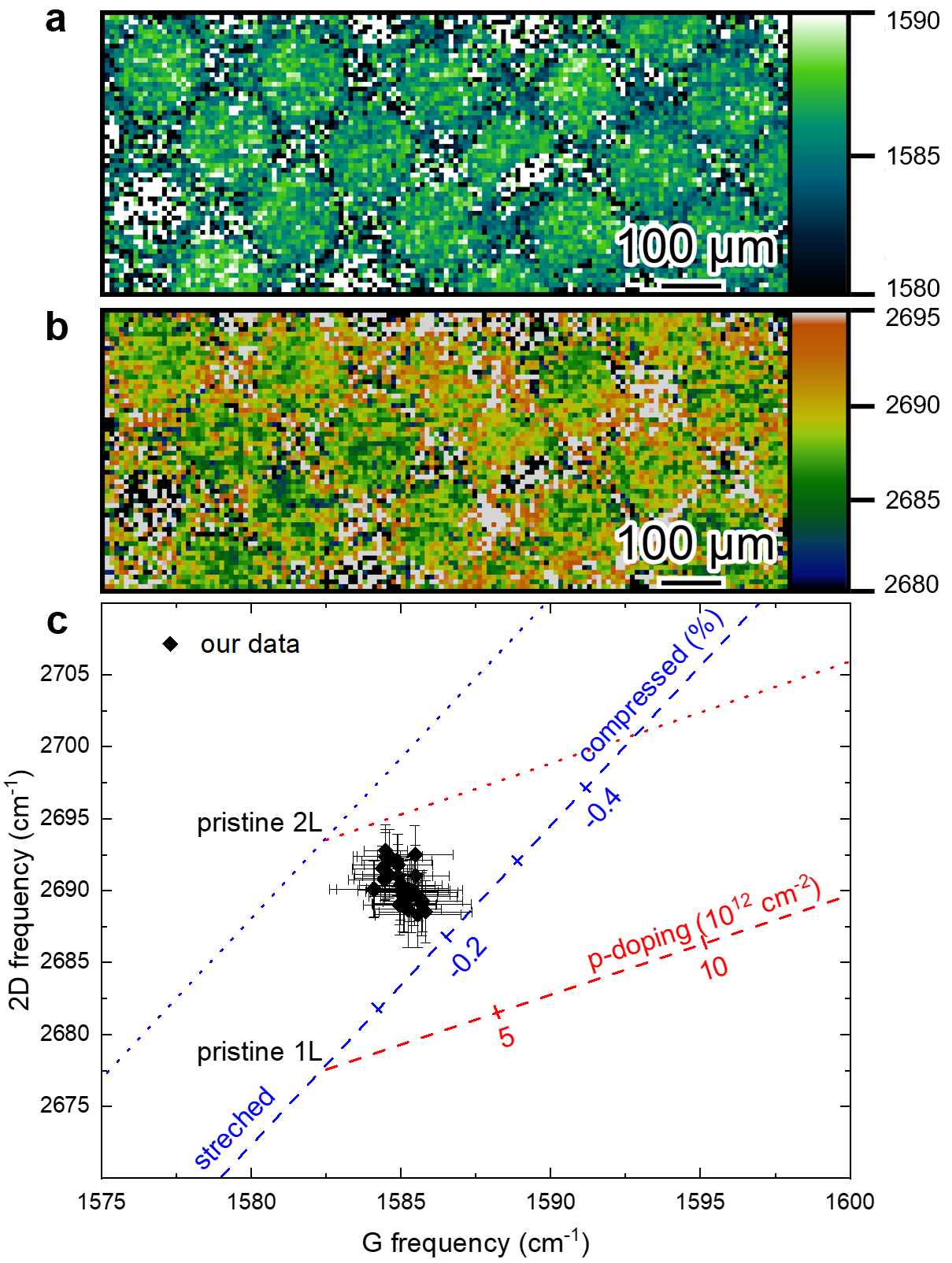}   
    \captionsetup{justification=raggedright,singlelinecheck=false}
 \caption{Raman maps over a larger sample area showing the peak positions of the $G$ mode \textbf{a} and the 2D mode \textbf{b}. $2D/G$ for triple layer graphene summarized in \textbf{c} as shown in \textbf{a} and \textbf{b} averaged over 276 data points for each hole. Dashed and dotted lines are for mono- and bilayer graphene, respectively, and are taken from ref.~\cite{Kim.2016}.}
 \label{Raman}
\end{figure}

Single-point spectra (not shown here)  of our triple layer samples resemble spectra of monolayer graphene and thus do not show the typical signature of a bilayer graphene. In order to get better statistics and to compare different holes, we utilized the extended area mapping feature of a Renishaw InVia$\mathrm{^{TM}}$ confocal Raman microscope. In Fig.~\ref{Raman}~\textbf{a} and \textbf{b}, such maps featuring 14 holes are presented, showcasing both the position of the $G$ mode and that of the $2D$ mode. In case of the $G$ mode, the holes can be clearly distinguished from the surrounding Si substrate, while in the case of the $2D$ mode, the holes appear more blurry. 

From each hole, 276 data points were collected, with a step size of 8~$\mathrm{\text\textmu m}$ per data point, and are summarized via the software package Gwyddion~\cite{Necas.2012} yielding a probability distribution of the mode positions which is then fitted with a Gaussian function. Thus, each data point in Fig.~\ref{Raman}~\textbf{c} corresponds to an individual hole, with the  error bars indicating the standard deviation for the corresponding average value of this particular hole. To this plot, we have added the linear functions (dashed lines) established by Lee \textit{et al}.~\cite{Lee.2012} representing the shift of the Raman modes of monolayer graphene in the presence of strain (blue) and charge carriers doping (red). The position of the Raman modes for pristine monolayer graphene is given by the intersection. For bilayer graphene, the respective data is shifted~\cite{Kim.2016} and presented here by the dotted lines. 

As Fig.~\ref{Raman}~\textbf{c} shows, we find constant strain and doping levels within each hole and a remarkable reproducibility across the different holes. Notably, our data from triple layer graphene lies in between the values for mono- and bilayer graphene indicating a decoupling of the respective layers, as also shown in the single-point spectra indicating a monolayer of graphene. This makes a straightforward quantification of our data difficult. If we take bilayer graphene as a reference~\cite{Kim.2016} for quantification, averaging over all our data points and using the vector model of Lee \textit{et al}., we would arrive at hole doping with (5.5$\thinspace \pm \thinspace$1.5)$\thinspace \times \thinspace$10$^{12}$~cm$^{-2}$ and tensile strain of (0.2$\thinspace \pm \thinspace$0.09)\thinspace\%. While these numbers might give a first indication that the triple layers are slightly strained and $p$-doped, we chose a complementary approach to gain further insight into the electronic properties of our graphene samples. 

\subsection{THz spectroscopy: Mobility and charge carrier density}
\begin{figure}[ht]
    \centering
    \includegraphics[width = 0.5\textwidth]{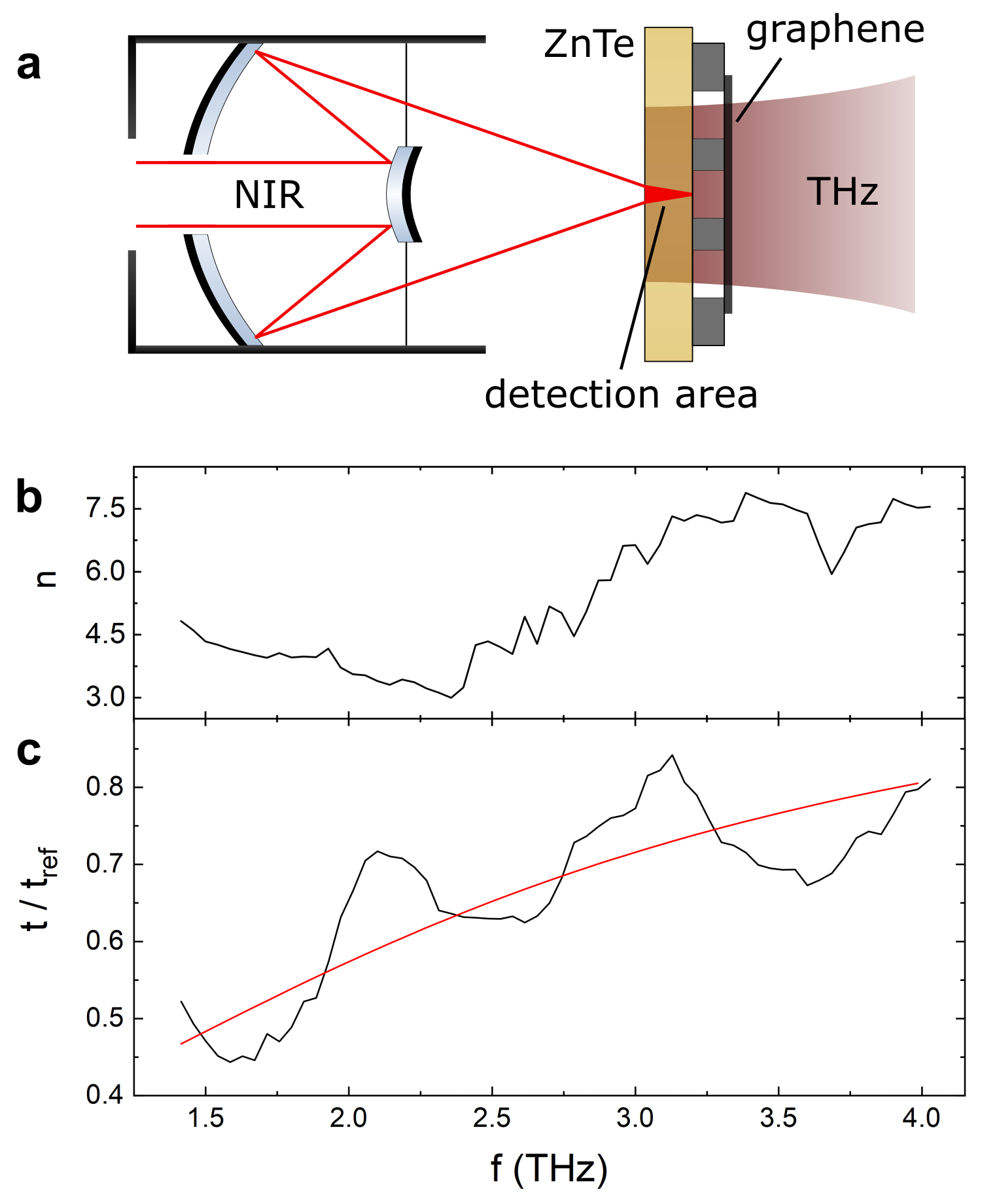}  
    \captionsetup{justification=raggedright,singlelinecheck=false}
    \caption{\textbf{a} Sketch of the all-optical near-field microscope to measure the THz transmission through the subwavelength sized holes. The THz pulse penetrates the graphene layer before it passes through the metallic hole. The transmitted THz field is measured in the ZnTe crystal that is in direct contact with the sample holder. \textbf{b} Measured refractive index of the holes in the steel sample holder. \textbf{c} Averaged THz transmission (black line) through the graphene sample. The influence of the holes is taken into account through the reference measurement. The oscillations are caused by multiple reflections in the ZnTe crystal. The red line represents a fit to the experimental data to extract the charge carrier density and mobility.}
\label{THz}
\end{figure}

Terahertz (THz) spectroscopy is a versatile tool to investigate the carrier density and mobility without the need of electrical contacts~\cite{LloydHughes.2012}, the processing of which would again contaminate our samples. In particular, in thin conductive films like graphene, the interaction with THz radiation is dominated by intraband absorption~\cite{Dawlaty.2008}. Due to the large wavelength of 300~$\mathrm{\text\textmu m}$ at 1~THz photon frequency, the spot size of conventional THz spectrometers is on the order of about 1~mm and thus requires large samples. To overcome this obstacle, we performed the THz measurements with an all optical near-field microscope~\cite{Adam.2011} by bringing our graphene samples in close proximity to an electro-optical crystal (ZnTe, compare Fig.~\ref{THz}~\textbf{a}. Note, that for these experiments we had to use the steel substrates with holes in it, due to the fact that the quality of the Si hole edges was not sufficient enough. A near-infrared beam serves as probe to measure the THz transient. 

The spatial resolution is in this case mostly dependent on the numerical aperture of the near-infrared beam and the thickness of the electro-optical crystal. With a 150~µm thick ZnTe crystal, we can resolve objects smaller than 100~$\mathrm{\text\textmu m}$. A sketch of the near-field optics is shown in Fig.~\ref{THz}~\textbf{a}. To analyze the data with a thin film model, the refractive index of the substrate needs to be known. This was measured by comparing the transmission through one of the uncovered holes with the impinging THz field, the result is plotted in Fig.~\ref{THz}~\textbf{b}. 

The freestanding graphene triple layer was then measured by comparing the transmission through a hole covered by graphene with a reference measurement through a bare hole. As the optical properties of the ZnTe crystal can vary spatially due to defects, we performed reference measurements for each of the sample measurements through neighboring holes, the resulting transmission spectra were averaged to reduce noise. The spectrum of the transmitted electric field is shown in Fig.~\ref{THz}~\textbf{c}. The oscillations that can be observed in the spectrum are caused by multiple reflections in the ZnTe crystal and cannot be entirely avoided due to the dispersion in the metallic hole.

A thin film model~\cite{Glover.1957} in combination with Drude conductivity was fitted to the experimental data (red line in Fig.~\ref{THz}~\textbf{c}). Assuming three layers of graphene, we derive the charge carrier density in each of the layers to be (2.1$\thinspace \pm \thinspace$0.2)$\thinspace\times\thinspace$10$^{13}$~cm$^{-2}$ with a mobility of (4.3$\thinspace \pm \thinspace$2.4)$\thinspace\times\thinspace$10$^3$~cm$^2$V$^{-1}$s$^{-1}$. Performing the data analysis for individual holes, the variation between the holes were smaller than the error, confirming the homogeneity of the freestanding graphene film. The relatively large error for the mobility is related to the dispersion of the metallic hole: as the fitting did not give meaningful results considering the frequency dependence of the refractive index, we performed the analysis with the averaged value of the measured range. The observed high mobility is well known for free standing graphene films~\cite{Bolotin.2008}. The charge carrier density is also found to be high and in good agreement with the doping level reported previously for graphene prepared in this way due to the functionalization with acetate groups~\cite{Madau.2020}.

\subsection{Ion-Electron Conversion}
\begin{figure}[htb]
   \centering
   \includegraphics[width=0.5\textwidth]{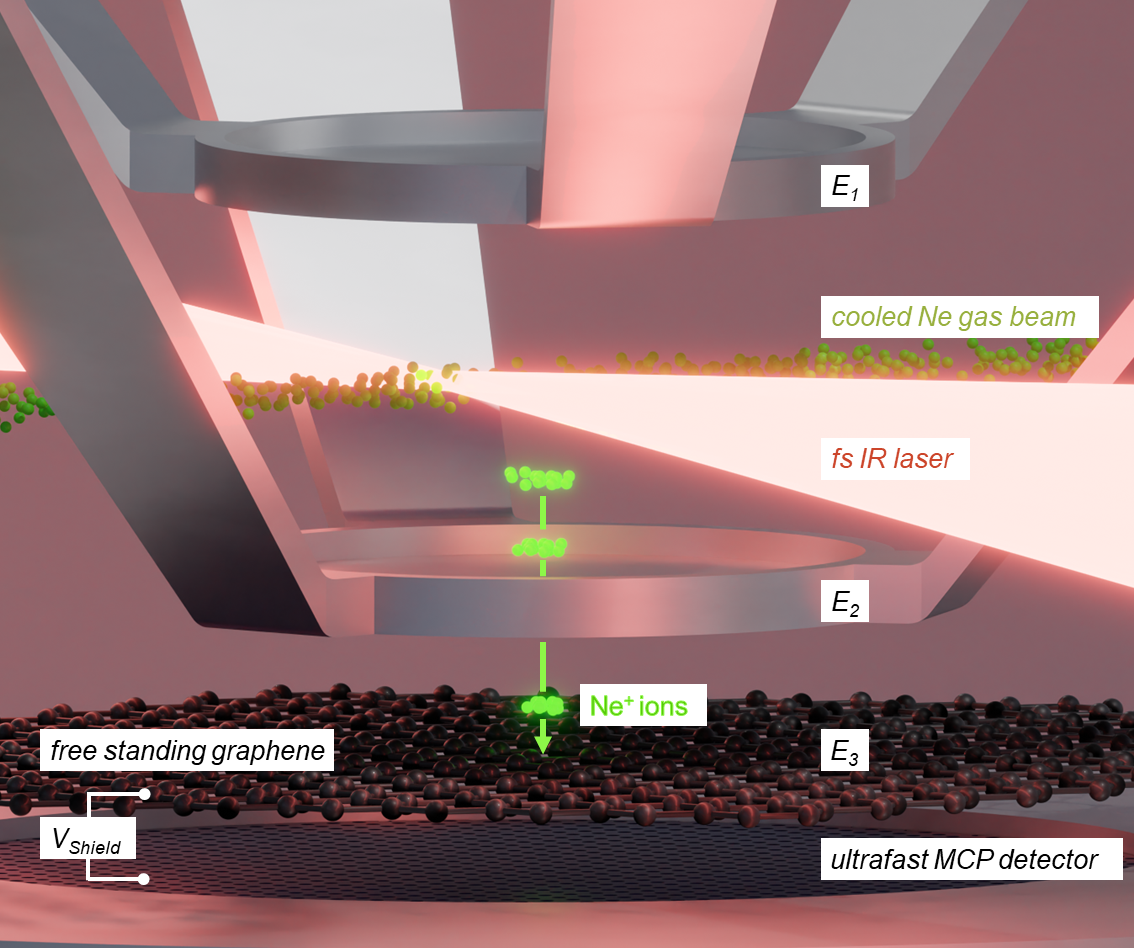}
   \captionsetup{justification=raggedright,singlelinecheck=false}
   \caption{3D rendering of the setup for the ion-electron conversion experiments. A fs-IR-laser (red) is focused between two electrodes $E\mathrm{_1}$ and $E\mathrm{_2}$. The laser intersects a supersonic geometrically cooled neon gas jet (green particles), creating ions which are accelerated by an electric field towards $E\mathrm{_2}$. Ions are extracted though a pinhole of 80~$\mathrm{\text\textmu m}$ diameter in the center of $E\mathrm{_2}$ and move towards the graphene target ($E\mathrm{_3}$) above an ultra-fast MCP detector. The height and tilt angle of each electrode and the target can be independently adjusted relative to the cooled gas jet to ensure optimum alignment. A voltage V$_{Shield}$ can be applied between the graphene and the front of the detector to accelerate emitted secondary electrons onto the MCP.}
   \label{Buncher}
\end{figure}

Finally, we address the suitability of our samples to serve as ion-electron converters to be used for example in a ToF-experiment. To this end we implemented the graphene samples into our recently developed ps ion source. The set-up is described in detail in refs.~\cite{Golombek.2021,Kalkhoff.2023} and will thus be discussed here only briefly: Monoenergetic keV ion pulses are generated by focusing an infrared (IR) femtosecond laser pulse into an ion buncher (see Fig.~\ref{Buncher}). Tunnel ionization of neutral gas atoms entrained in a supersonic gas jet is induced by a high intensity fs laser pulse focused between the two electrodes $E_1$ and $E_2$. These will accelerate the photo-ionized ions towards an 80~$\mathrm{\text\textmu m}$ extraction pinhole in the bottom electrode $E\mathrm{_2}$. With this approach we arrive at ion pulses as short as 18~ps full width at half maximum (FWHM)~\cite{Kalkhoff.2023}. 

The graphene targets are mounted 1.1~mm above an ultrafast multichannel plate (MCP) detector (Photonis Gen2 ToF Detector) as depicted in Fig.~\ref{Buncher}. Just like the electrodes $E\mathrm{_1}$ and $E\mathrm{_2}$, the graphene substrate can be biased with (0--210)~kV with regard to $E_\mathrm{_2}$. Such a three-electrode setup is known as Wiley-McLaren configuration~\cite{Wiley.1955}, where the graphene target acts as the third electrode, $E_\mathrm{_3}$. Any ion generated in the ionization volume is accelerated towards the graphene surface. Particles emitted from, or transmitted through, graphene due to ion impacts are readily detected by the MCP. 

In addition, the front of the MCP can be biased with respect to the graphene target to de- and accelerate particles emitted from the graphene towards the detector. This allows for mass-spectrometry by a time-of-flight analysis as any potential applied to accelerate (decelerate) electrons will inherently result in the mass/charge-dependent deceleration (acceleration) of ions passing through the freestanding graphene target. The potential difference between the freestanding graphene and the MCP front is termed as V$_{Shield}$ and if $e\cdot V_{Shield}$ is lower than the kinetic energy E$_{kin}$ of the ions, the resulting spectrum will include any emitted electrons and ions. 

\begin{figure}[htb]
    \centering
   \includegraphics[width=0.5\textwidth]{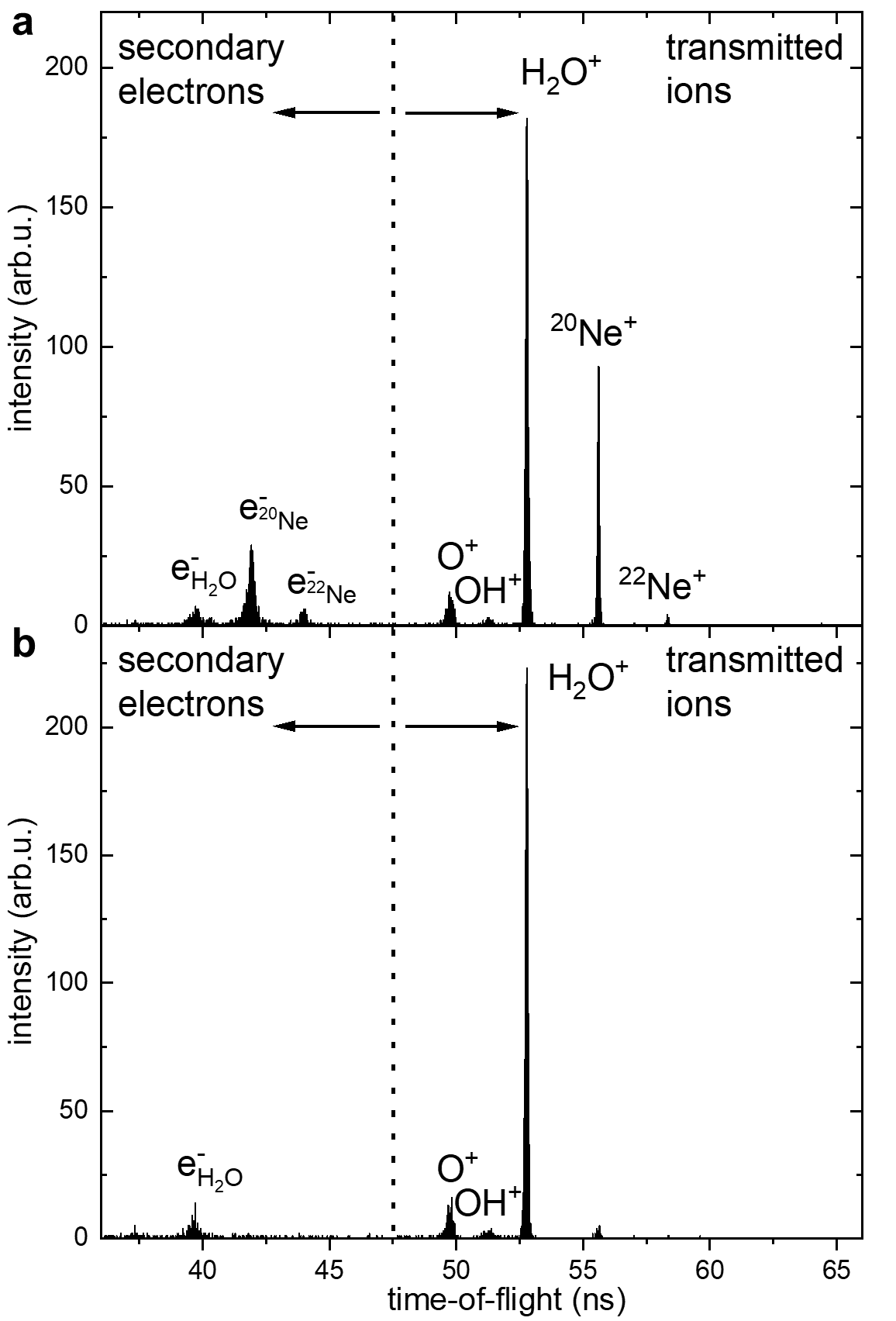}
   \captionsetup{justification=raggedright,singlelinecheck=false}
    \caption{Time-of-flight spectrum of emitted particles from the graphene target onto a MCP surface after bombardment with ions, \textbf{a}~while the Ne gas jet active and \textbf{b}~without the Ne gas jet. Primary ions are accelerated towards the graphene target with 3.5~keV and deaccelerated by a 2~kV electric field after the impact with the target. Events left of the dashed line are can be assigned to secondary electrons emitted from the graphene, events right of the dashed line are transmitted ions with a remaining energy of 1.5~keV. $\mathrm{H_2 O}$$^+$,OH$^+$ and O$^+$ can be contributed to the residual gas in the UHV chamber.}
   \label{Conversion}
\end{figure}

To measure the flight time of particles emitted from the graphene target, the output signal of the MCP is fed into an ultrafast oscilloscope (LeCroy WavePro404). The arrival time of each MCP pulse is measured against a signal of a fast photodiode which is triggered by a small fraction of the ionizing laser pulse. The time difference between the 50\thinspace\% mark of the photodiode signal and 50\thinspace\% mark of the MCP signal are recorded in a histogram resulting in the time-of-flight spectra seen in Fig.~\ref{Conversion}.

The spectrum shown in Fig.~\ref{Conversion}~\textbf{a} was collected from ions which were first accelerated to a kinetic energy of E$_{kin}$\thinspace=\thinspace3.5~keV and then after passing through the graphene target decelerated by e$\cdot$V$_{Shield}$$\thinspace=\thinspace$2.0~keV. Since the ions hit the MCP with a remaining energy of E$^{'}_{kin}$\thinspace=\thinspace1.5~keV, they still result in a detectable signal, making the \textit{transmitted} ions observable. If the neon gas jet is on, as shown in Fig.~\ref{Conversion}~\textbf{a}, peaks become clearly visible in the spectrum, which we therefore attribute to $\mathrm{^{20}}$Ne$^+$ and its isotope $\mathrm{^{22}}$Ne$^+$. When the gas jet is turned off, see Fig.~\ref{Conversion}~\textbf{b}, the signal nearly vanishes (not completely as there is a small leakage of the nozzle producing the neon gas jet). 

The arrival time of said signals can be converted to a mass-to-charge ratio, and with the assumption that the main signal indeed stems from $\mathrm{^{20}}$Ne$^+$ and $\mathrm{^{22}}$Ne$^+$, respectively, the ToF signal of other molecules and species can be determined. The signal at 52.7~ns can then be attributed to $\mathrm{H_2 O^+}$ from the residual gas, and 49.7~ns correspond to O, as well as OH at 51.2~ns, stemming from fragmented $\mathrm{H_2 O}$ due to the ionization process. Signals with even shorter arrival times can be explained by secondary electrons emitted by the graphene target after an ion impact. One indicator for this is that only certain peaks disappear when the neon jet is absent. Therefore these must result from the gas jet itself and can only be explained by secondary electrons emitted from the graphene target after an ion impact. 

To calculate the electron yield we look at the total numbers of events which can be assigned to a secondary electron and the corresponding transmitted ion.
For Ne$^+$, we measure that 80\thinspace\% of ions will lead to an emission of an electron. But due to the fact that electrons at 2.0~keV~\cite{Galanti.1971,Blase.2017} and ions at 1.5~keV~\cite{Brehm.1995} do not have the same detection efficiency, we over-estimate the total number of electrons and under-estimate the number of ions. This brings our results closer to the 0.5 electrons per ion that theoretical work by Yao \textit{et al}. predicts~\cite{Y.YaoA.MetzlaffL.BreuerandA.Schleife.2023}.  


\section{\label{Conclusion} Conclusion and outlook \protect}
In conclusion, we have demonstrated the manufacturing of samples with graphene triple layers spanning circular holes of 150~$\mathrm{\text\textmu m}$ without the use of any polymer. Our method has a high coverage ($>$90\thinspace\%) and results in clean, flat, homogeneously doped and strained samples. Optical profilometry has in this context proven to be an efficient tool for rapid quality control, ensuring high throughput. By means of THz spectroscopy we could show that our samples feature a high charge carrier density and mobility. This together with the phonon modes confirm that the triple layer targets are structurally and electronically very similar to monolayer-graphene, facilitating the comparison of experimental data with theoretical predictions.   

In addition, we have shown that our triple layer targets are efficient ion-to-electron converters and can be applied for ToF mass spectrometry. We find them easy to handle and to be sufficiently robust against irradiation damage. In comparison to previously used carbon foils they are ultrathin and their thickness is defined with an accuracy of ca.~1\AA~which makes them particularly interesting for fundamental studies in the field of ion-solid interaction.  With respect to applications, our graphene samples could be used, e.g., to greatly improve the mass resolution of ToF spectrometers. Conventional set-ups use carbon foils to generate the START signal, the timing of which suffers from angular and energy straggling, which would be minimized or even absent (no collision cascade) in as-thin-as-it-gets films. 

For this work, the overall dimensions of the targets has been specifically adapted to fit into our ps-ion source but the manufacturing process can be up-scaled to even larger hole substrates. Our method can be used to fabricate graphene targets for future pump-probe-experiments with ultrashort ion pulses but because of their good graphene/substrate area ratio of 58\thinspace\%, they are also applicable for stripping purposes and any type of coincidence measurements. Due to their unique properties, we further expect our targets to be useful in other applications such as pressure sensors or in the field of NEMS and MEMS, respectively.

\begin{acknowledgments}
The authors are greatly indebted to Ping Zhou for his assistance with the laser system. We thank the Deutsche Forschungsgemeinschaft (DFG) for their financial support of project C05 and B09 within the Collaborative Research Center (CRC) 1242 ‘Non-equilibrium dynamics in the time domain’ (project number 278162697) and the Federal Ministry of Education and Research (BMBF) for funding within project 05K16PG1.
\end{acknowledgments}

\nocite{*}
\bibliography{aipsamp}

\begin{thebibliography}{56}%
\makeatletter
\providecommand \@ifxundefined [1]{%
 \@ifx{#1\undefined}
}%
\providecommand \@ifnum [1]{%
 \ifnum #1\expandafter \@firstoftwo
 \else \expandafter \@secondoftwo
 \fi
}%
\providecommand \@ifx [1]{%
 \ifx #1\expandafter \@firstoftwo
 \else \expandafter \@secondoftwo
 \fi
}%
\providecommand \natexlab [1]{#1}%
\providecommand \enquote  [1]{``#1''}%
\providecommand \bibnamefont  [1]{#1}%
\providecommand \bibfnamefont [1]{#1}%
\providecommand \citenamefont [1]{#1}%
\providecommand \href@noop [0]{\@secondoftwo}%
\providecommand \href [0]{\begingroup \@sanitize@url \@href}%
\providecommand \@href[1]{\@@startlink{#1}\@@href}%
\providecommand \@@href[1]{\endgroup#1\@@endlink}%
\providecommand \@sanitize@url [0]{\catcode `\\12\catcode `\$12\catcode `\&12\catcode `\#12\catcode `\^12\catcode `\_12\catcode `\%12\relax}%
\providecommand \@@startlink[1]{}%
\providecommand \@@endlink[0]{}%
\providecommand \url  [0]{\begingroup\@sanitize@url \@url }%
\providecommand \@url [1]{\endgroup\@href {#1}{\urlprefix }}%
\providecommand \urlprefix  [0]{URL }%
\providecommand \Eprint [0]{\href }%
\providecommand \doibase [0]{https://doi.org/}%
\providecommand \selectlanguage [0]{\@gobble}%
\providecommand \bibinfo  [0]{\@secondoftwo}%
\providecommand \bibfield  [0]{\@secondoftwo}%
\providecommand \translation [1]{[#1]}%
\providecommand \BibitemOpen [0]{}%
\providecommand \bibitemStop [0]{}%
\providecommand \bibitemNoStop [0]{.\EOS\space}%
\providecommand \EOS [0]{\spacefactor3000\relax}%
\providecommand \BibitemShut  [1]{\csname bibitem#1\endcsname}%
\let\auto@bib@innerbib\@empty
\bibitem [{\citenamefont {Li}\ and\ \citenamefont {Zhou}(2012)}]{Li.2012}%
  \BibitemOpen
  \bibfield  {author} {\bibinfo {author} {\bibfnamefont {H.}~\bibnamefont {Li}}\ and\ \bibinfo {author} {\bibfnamefont {H.}~\bibnamefont {Zhou}},\ }\bibfield  {title} {\bibinfo {title} {Enhancing the performances of {L}i-ion batteries by carbon-coating: present and future},\ }\href {https://doi.org/10.1039/C1CC14764A} {\bibfield  {journal} {\bibinfo  {journal} {Chemical Communications}\ }\textbf {\bibinfo {volume} {48}},\ \bibinfo {pages} {1201} (\bibinfo {year} {2012})}\BibitemShut {NoStop}%
\bibitem [{\citenamefont {Adeel}\ \emph {et~al.}(2022)\citenamefont {Adeel}, \citenamefont {Asif}, \citenamefont {Canzonieri}, \citenamefont {Barai}, \citenamefont {Rahman}, \citenamefont {Daniele},\ and\ \citenamefont {Rizzolio}}]{Adeel.2022}%
  \BibitemOpen
  \bibfield  {author} {\bibinfo {author} {\bibfnamefont {M.}~\bibnamefont {Adeel}}, \bibinfo {author} {\bibfnamefont {K.}~\bibnamefont {Asif}}, \bibinfo {author} {\bibfnamefont {V.}~\bibnamefont {Canzonieri}}, \bibinfo {author} {\bibfnamefont {H.~R.}\ \bibnamefont {Barai}}, \bibinfo {author} {\bibfnamefont {M.~M.}\ \bibnamefont {Rahman}}, \bibinfo {author} {\bibfnamefont {S.}~\bibnamefont {Daniele}},\ and\ \bibinfo {author} {\bibfnamefont {F.}~\bibnamefont {Rizzolio}},\ }\bibfield  {title} {\bibinfo {title} {Controlled, partially exfoliated, self-supported functionalized flexible graphitic carbon foil for ultrasensitive detection of sars-cov-2 spike protein},\ }\href {https://doi.org/10.1016/j.snb.2022.131591} {\bibfield  {journal} {\bibinfo  {journal} {Sensors and Actuators B: Chemical}\ }\textbf {\bibinfo {volume} {359}},\ \bibinfo {pages} {131591} (\bibinfo {year} {2022})}\BibitemShut {NoStop}%
\bibitem [{\citenamefont {Oliphant}(1930)}]{Oliphant.1930b}%
  \BibitemOpen
  \bibfield  {author} {\bibinfo {author} {\bibfnamefont {M.}~\bibnamefont {Oliphant}},\ }\bibfield  {title} {\bibinfo {title} {The liberation of electrons from metal surfaces by positive ions. part {I}.---experimental},\ }\href {https://doi.org/10.1098/rspa.1930.0065} {\bibfield  {journal} {\bibinfo  {journal} {Proceedings of the Royal Society of London. Series A, Containing Papers of a Mathematical and Physical Character}\ }\textbf {\bibinfo {volume} {127}},\ \bibinfo {pages} {373} (\bibinfo {year} {1930})}\BibitemShut {NoStop}%
\bibitem [{\citenamefont {Oliphant}\ and\ \citenamefont {Moon}(1930)}]{Oliphant.1930}%
  \BibitemOpen
  \bibfield  {author} {\bibinfo {author} {\bibfnamefont {M.}~\bibnamefont {Oliphant}}\ and\ \bibinfo {author} {\bibfnamefont {P.}~\bibnamefont {Moon}},\ }\bibfield  {title} {\bibinfo {title} {The liberation of electrons from metal surfaces by positive ions. part {II}.---theoretical},\ }\href {https://doi.org/10.1098/rspa.1930.0066} {\bibfield  {journal} {\bibinfo  {journal} {Proceedings of the Royal Society of London. Series A, Containing Papers of a Mathematical and Physical Character}\ }\textbf {\bibinfo {volume} {127}},\ \bibinfo {pages} {388} (\bibinfo {year} {1930})}\BibitemShut {NoStop}%
\bibitem [{\citenamefont {Ebert}\ \emph {et~al.}(2014)\citenamefont {Ebert}, \citenamefont {Allegrini}, \citenamefont {Fuselier}, \citenamefont {Nicolaou}, \citenamefont {Bedworth}, \citenamefont {Sinton},\ and\ \citenamefont {Trattner}}]{Ebert.2014}%
  \BibitemOpen
  \bibfield  {author} {\bibinfo {author} {\bibfnamefont {R.~W.}\ \bibnamefont {Ebert}}, \bibinfo {author} {\bibfnamefont {F.}~\bibnamefont {Allegrini}}, \bibinfo {author} {\bibfnamefont {S.~A.}\ \bibnamefont {Fuselier}}, \bibinfo {author} {\bibfnamefont {G.}~\bibnamefont {Nicolaou}}, \bibinfo {author} {\bibfnamefont {P.}~\bibnamefont {Bedworth}}, \bibinfo {author} {\bibfnamefont {S.}~\bibnamefont {Sinton}},\ and\ \bibinfo {author} {\bibfnamefont {K.~J.}\ \bibnamefont {Trattner}},\ }\bibfield  {title} {\bibinfo {title} {Angular scattering of 1-50 ke{V} ions through graphene and thin carbon foils: potential applications for space plasma instrumentation},\ }\href {https://doi.org/10.1063/1.4866850} {\bibfield  {journal} {\bibinfo  {journal} {Review of Scientific Instruments}\ }\textbf {\bibinfo {volume} {85}},\ \bibinfo {pages} {033302} (\bibinfo {year} {2014})}\BibitemShut {NoStop}%
\bibitem [{\citenamefont {Allegrini}\ \emph {et~al.}(2003)\citenamefont {Allegrini}, \citenamefont {Wimmer-Schweingruber}, \citenamefont {Wurz},\ and\ \citenamefont {Bochsler}}]{Allegrini.2003}%
  \BibitemOpen
  \bibfield  {author} {\bibinfo {author} {\bibfnamefont {F.}~\bibnamefont {Allegrini}}, \bibinfo {author} {\bibfnamefont {R.~F.}\ \bibnamefont {Wimmer-Schweingruber}}, \bibinfo {author} {\bibfnamefont {P.}~\bibnamefont {Wurz}},\ and\ \bibinfo {author} {\bibfnamefont {P.}~\bibnamefont {Bochsler}},\ }\bibfield  {title} {\bibinfo {title} {Determination of low-energy ion-induced electron yields from thin carbon foils},\ }\href {https://doi.org/10.1016/S0168-583X(03)01705-1} {\bibfield  {journal} {\bibinfo  {journal} {Nuclear Instruments and Methods in Physics Research Section B: Beam Interactions with Materials and Atoms}\ }\textbf {\bibinfo {volume} {211}},\ \bibinfo {pages} {487} (\bibinfo {year} {2003})}\BibitemShut {NoStop}%
\bibitem [{\citenamefont {Allegrini}\ \emph {et~al.}(2016)\citenamefont {Allegrini}, \citenamefont {Ebert},\ and\ \citenamefont {Funsten}}]{Allegrini.2016}%
  \BibitemOpen
  \bibfield  {author} {\bibinfo {author} {\bibfnamefont {F.}~\bibnamefont {Allegrini}}, \bibinfo {author} {\bibfnamefont {R.~W.}\ \bibnamefont {Ebert}},\ and\ \bibinfo {author} {\bibfnamefont {H.~O.}\ \bibnamefont {Funsten}},\ }\bibfield  {title} {\bibinfo {title} {Carbon foils for space plasma instrumentation},\ }\href {https://doi.org/10.1002/2016JA022570} {\bibfield  {journal} {\bibinfo  {journal} {Journal of Geophysical Research: Space Physics}\ }\textbf {\bibinfo {volume} {121}},\ \bibinfo {pages} {3931} (\bibinfo {year} {2016})}\BibitemShut {NoStop}%
\bibitem [{\citenamefont {Vira}\ \emph {et~al.}(2020)\citenamefont {Vira}, \citenamefont {Fernandes}, \citenamefont {Skoug}, \citenamefont {Funsten},\ and\ \citenamefont {Reisenfeld}}]{Vira.2020}%
  \BibitemOpen
  \bibfield  {author} {\bibinfo {author} {\bibfnamefont {A.~D.}\ \bibnamefont {Vira}}, \bibinfo {author} {\bibfnamefont {P.~A.}\ \bibnamefont {Fernandes}}, \bibinfo {author} {\bibfnamefont {R.~M.}\ \bibnamefont {Skoug}}, \bibinfo {author} {\bibfnamefont {H.~O.}\ \bibnamefont {Funsten}},\ and\ \bibinfo {author} {\bibfnamefont {D.~B.}\ \bibnamefont {Reisenfeld}},\ }\bibfield  {title} {\bibinfo {title} {Understanding mass resolution of foil--based time--of--flight mass spectrometry},\ }\href {https://doi.org/10.1029/2020JA027971} {\bibfield  {journal} {\bibinfo  {journal} {Journal of Geophysical Research: Space Physics}\ }\textbf {\bibinfo {volume} {125}},\ \bibinfo {pages} {e2020JA027971} (\bibinfo {year} {2020})}\BibitemShut {NoStop}%
\bibitem [{\citenamefont {Menendez}\ \emph {et~al.}(1986)\citenamefont {Menendez}, \citenamefont {Duncan}, \citenamefont {Berry}, \citenamefont {Sellin}, \citenamefont {Meckbach}, \citenamefont {Focke},\ and\ \citenamefont {Nemirovsky}}]{Menendez.1986}%
  \BibitemOpen
  \bibfield  {author} {\bibinfo {author} {\bibfnamefont {M.~G.}\ \bibnamefont {Menendez}}, \bibinfo {author} {\bibfnamefont {M.~M.}\ \bibnamefont {Duncan}}, \bibinfo {author} {\bibfnamefont {S.~D.}\ \bibnamefont {Berry}}, \bibinfo {author} {\bibfnamefont {I.~A.}\ \bibnamefont {Sellin}}, \bibinfo {author} {\bibfnamefont {W.}~\bibnamefont {Meckbach}}, \bibinfo {author} {\bibfnamefont {P.}~\bibnamefont {Focke}},\ and\ \bibinfo {author} {\bibfnamefont {I.~B.}\ \bibnamefont {Nemirovsky}},\ }\bibfield  {title} {\bibinfo {title} {Coincidence experiment concerning the origin of convoy electrons produced by swift deuterium beams traversing carbon foils},\ }\href {https://doi.org/10.1103/PhysRevA.33.2160} {\bibfield  {journal} {\bibinfo  {journal} {Physical review. A, General physics}\ }\textbf {\bibinfo {volume} {33}},\ \bibinfo {pages} {2160} (\bibinfo {year} {1986})}\BibitemShut {NoStop}%
\bibitem [{\citenamefont {McComas}\ and\ \citenamefont {Nordholt}(1990)}]{McComas.1990}%
  \BibitemOpen
  \bibfield  {author} {\bibinfo {author} {\bibfnamefont {D.~J.}\ \bibnamefont {McComas}}\ and\ \bibinfo {author} {\bibfnamefont {J.~E.}\ \bibnamefont {Nordholt}},\ }\bibfield  {title} {\bibinfo {title} {New approach to 3-d, high sensitivity, high mass resolution space plasma composition measurements},\ }\href {https://doi.org/10.1063/1.1141692} {\bibfield  {journal} {\bibinfo  {journal} {Review of Scientific Instruments}\ }\textbf {\bibinfo {volume} {61}},\ \bibinfo {pages} {3095} (\bibinfo {year} {1990})}\BibitemShut {NoStop}%
\bibitem [{\citenamefont {Hole{\v{n}}{\'a}k}\ \emph {et~al.}(2021)\citenamefont {Hole{\v{n}}{\'a}k}, \citenamefont {Lohmann}, \citenamefont {Sekula},\ and\ \citenamefont {Primetzhofer}}]{Holenak.2021}%
  \BibitemOpen
  \bibfield  {author} {\bibinfo {author} {\bibfnamefont {R.}~\bibnamefont {Hole{\v{n}}{\'a}k}}, \bibinfo {author} {\bibfnamefont {S.}~\bibnamefont {Lohmann}}, \bibinfo {author} {\bibfnamefont {F.}~\bibnamefont {Sekula}},\ and\ \bibinfo {author} {\bibfnamefont {D.}~\bibnamefont {Primetzhofer}},\ }\bibfield  {title} {\bibinfo {title} {Simultaneous assessment of energy, charge state and angular distribution for medium energy ions interacting with ultra-thin self-supporting targets: A time-of-flight approach},\ }\href {https://doi.org/10.1016/j.vacuum.2020.109988} {\bibfield  {journal} {\bibinfo  {journal} {Vacuum}\ }\textbf {\bibinfo {volume} {185}},\ \bibinfo {pages} {109988} (\bibinfo {year} {2021})}\BibitemShut {NoStop}%
\bibitem [{\citenamefont {Moak}(1976)}]{Moak.1976}%
  \BibitemOpen
  \bibfield  {author} {\bibinfo {author} {\bibfnamefont {C.~D.}\ \bibnamefont {Moak}},\ }\bibfield  {title} {\bibinfo {title} {Stripping in foils and gases},\ }\href {https://doi.org/10.1109/TNS.1976.4328419} {\bibfield  {journal} {\bibinfo  {journal} {IEEE Transactions on Nuclear Science}\ }\textbf {\bibinfo {volume} {23}},\ \bibinfo {pages} {1126} (\bibinfo {year} {1976})}\BibitemShut {NoStop}%
\bibitem [{\citenamefont {Hattass}\ \emph {et~al.}(1999)\citenamefont {Hattass}, \citenamefont {Schenkel}, \citenamefont {Hamza}, \citenamefont {Barnes}, \citenamefont {Newman}, \citenamefont {McDonald}, \citenamefont {Niedermayr}, \citenamefont {Machicoane},\ and\ \citenamefont {Schneider}}]{Hattass.1999}%
  \BibitemOpen
  \bibfield  {author} {\bibinfo {author} {\bibfnamefont {M.}~\bibnamefont {Hattass}}, \bibinfo {author} {\bibfnamefont {T.}~\bibnamefont {Schenkel}}, \bibinfo {author} {\bibfnamefont {A.~V.}\ \bibnamefont {Hamza}}, \bibinfo {author} {\bibfnamefont {A.~V.}\ \bibnamefont {Barnes}}, \bibinfo {author} {\bibfnamefont {M.~W.}\ \bibnamefont {Newman}}, \bibinfo {author} {\bibfnamefont {J.~W.}\ \bibnamefont {McDonald}}, \bibinfo {author} {\bibfnamefont {T.~R.}\ \bibnamefont {Niedermayr}}, \bibinfo {author} {\bibfnamefont {G.~A.}\ \bibnamefont {Machicoane}},\ and\ \bibinfo {author} {\bibfnamefont {D.~H.}\ \bibnamefont {Schneider}},\ }\bibfield  {title} {\bibinfo {title} {Charge equilibration time of slow, highly charged ions in solids},\ }\href {https://doi.org/10.1103/PhysRevLett.82.4795} {\bibfield  {journal} {\bibinfo  {journal} {Physical Review Letters}\ }\textbf {\bibinfo {volume} {82}},\ \bibinfo {pages} {4795} (\bibinfo {year} {1999})}\BibitemShut {NoStop}%
\bibitem [{\citenamefont {Meyer}\ \emph {et~al.}(2007)\citenamefont {Meyer}, \citenamefont {Geim}, \citenamefont {Katsnelson}, \citenamefont {Novoselov}, \citenamefont {Booth},\ and\ \citenamefont {Roth}}]{Meyer.2007}%
  \BibitemOpen
  \bibfield  {author} {\bibinfo {author} {\bibfnamefont {J.~C.}\ \bibnamefont {Meyer}}, \bibinfo {author} {\bibfnamefont {A.~K.}\ \bibnamefont {Geim}}, \bibinfo {author} {\bibfnamefont {M.~I.}\ \bibnamefont {Katsnelson}}, \bibinfo {author} {\bibfnamefont {K.~S.}\ \bibnamefont {Novoselov}}, \bibinfo {author} {\bibfnamefont {T.~J.}\ \bibnamefont {Booth}},\ and\ \bibinfo {author} {\bibfnamefont {S.}~\bibnamefont {Roth}},\ }\bibfield  {title} {\bibinfo {title} {The structure of suspended graphene sheets},\ }\href {https://doi.org/10.1038/nature05545} {\bibfield  {journal} {\bibinfo  {journal} {Nature}\ }\textbf {\bibinfo {volume} {446}},\ \bibinfo {pages} {60} (\bibinfo {year} {2007})}\BibitemShut {NoStop}%
\bibitem [{\citenamefont {Chen}\ \emph {et~al.}(2016)\citenamefont {Chen}, \citenamefont {He}, \citenamefont {Huang}, \citenamefont {Huang}, \citenamefont {Shih}, \citenamefont {Chu}, \citenamefont {Kong}, \citenamefont {Li},\ and\ \citenamefont {Su}}]{Chen.2016}%
  \BibitemOpen
  \bibfield  {author} {\bibinfo {author} {\bibfnamefont {Y.-M.}\ \bibnamefont {Chen}}, \bibinfo {author} {\bibfnamefont {S.-M.}\ \bibnamefont {He}}, \bibinfo {author} {\bibfnamefont {C.-H.}\ \bibnamefont {Huang}}, \bibinfo {author} {\bibfnamefont {C.-C.}\ \bibnamefont {Huang}}, \bibinfo {author} {\bibfnamefont {W.-P.}\ \bibnamefont {Shih}}, \bibinfo {author} {\bibfnamefont {C.-L.}\ \bibnamefont {Chu}}, \bibinfo {author} {\bibfnamefont {J.}~\bibnamefont {Kong}}, \bibinfo {author} {\bibfnamefont {J.}~\bibnamefont {Li}},\ and\ \bibinfo {author} {\bibfnamefont {C.-Y.}\ \bibnamefont {Su}},\ }\bibfield  {title} {\bibinfo {title} {Ultra-large suspended graphene as a highly elastic membrane for capacitive pressure sensors},\ }\href {https://doi.org/10.1039/c5nr08668j} {\bibfield  {journal} {\bibinfo  {journal} {Nanoscale}\ }\textbf {\bibinfo {volume} {8}},\ \bibinfo {pages} {3555} (\bibinfo {year} {2016})}\BibitemShut {NoStop}%
\bibitem [{\citenamefont {Ros{\l}o{\'n}}\ \emph {et~al.}(2020)\citenamefont {Ros{\l}o{\'n}}, \citenamefont {Dolleman}, \citenamefont {Licona}, \citenamefont {Lee}, \citenamefont {{\v{S}}i{\v{s}}kins}, \citenamefont {Lebius}, \citenamefont {Madau{\ss}}, \citenamefont {Schleberger}, \citenamefont {Alijani}, \citenamefont {{van der Zant}},\ and\ \citenamefont {Steeneken}}]{Roson.2020}%
  \BibitemOpen
  \bibfield  {author} {\bibinfo {author} {\bibfnamefont {I.~E.}\ \bibnamefont {Ros{\l}o{\'n}}}, \bibinfo {author} {\bibfnamefont {R.~J.}\ \bibnamefont {Dolleman}}, \bibinfo {author} {\bibfnamefont {H.}~\bibnamefont {Licona}}, \bibinfo {author} {\bibfnamefont {M.}~\bibnamefont {Lee}}, \bibinfo {author} {\bibfnamefont {M.}~\bibnamefont {{\v{S}}i{\v{s}}kins}}, \bibinfo {author} {\bibfnamefont {H.}~\bibnamefont {Lebius}}, \bibinfo {author} {\bibfnamefont {L.}~\bibnamefont {Madau{\ss}}}, \bibinfo {author} {\bibfnamefont {M.}~\bibnamefont {Schleberger}}, \bibinfo {author} {\bibfnamefont {F.}~\bibnamefont {Alijani}}, \bibinfo {author} {\bibfnamefont {H.~S.~J.}\ \bibnamefont {{van der Zant}}},\ and\ \bibinfo {author} {\bibfnamefont {P.~G.}\ \bibnamefont {Steeneken}},\ }\bibfield  {title} {\bibinfo {title} {High-frequency gas effusion through nanopores in suspended graphene},\ }\href {https://doi.org/10.1038/s41467-020-19893-5} {\bibfield  {journal} {\bibinfo  {journal} {Nature Communications}\ }\textbf {\bibinfo
  {volume} {11}},\ \bibinfo {pages} {6025} (\bibinfo {year} {2020})}\BibitemShut {NoStop}%
\bibitem [{\citenamefont {Carvalho}\ \emph {et~al.}(2022)\citenamefont {Carvalho}, \citenamefont {Kulyk}, \citenamefont {Fernandes}, \citenamefont {Fortunato},\ and\ \citenamefont {Costa}}]{Carvalho.2022}%
  \BibitemOpen
  \bibfield  {author} {\bibinfo {author} {\bibfnamefont {A.~F.}\ \bibnamefont {Carvalho}}, \bibinfo {author} {\bibfnamefont {B.}~\bibnamefont {Kulyk}}, \bibinfo {author} {\bibfnamefont {A.~J.~S.}\ \bibnamefont {Fernandes}}, \bibinfo {author} {\bibfnamefont {E.}~\bibnamefont {Fortunato}},\ and\ \bibinfo {author} {\bibfnamefont {F.~M.}\ \bibnamefont {Costa}},\ }\bibfield  {title} {\bibinfo {title} {A review on the applications of graphene in mechanical transduction},\ }\href {https://doi.org/10.1002/adma.202101326} {\bibfield  {journal} {\bibinfo  {journal} {Advanced Materials}\ }\textbf {\bibinfo {volume} {34}},\ \bibinfo {pages} {e2101326} (\bibinfo {year} {2022})}\BibitemShut {NoStop}%
\bibitem [{\citenamefont {Romijn}\ \emph {et~al.}(2021)\citenamefont {Romijn}, \citenamefont {Dolleman}, \citenamefont {Singh}, \citenamefont {{van der Zant}}, \citenamefont {Steeneken}, \citenamefont {Sarro},\ and\ \citenamefont {Vollebregt}}]{Romijn.2021}%
  \BibitemOpen
  \bibfield  {author} {\bibinfo {author} {\bibfnamefont {J.}~\bibnamefont {Romijn}}, \bibinfo {author} {\bibfnamefont {R.~J.}\ \bibnamefont {Dolleman}}, \bibinfo {author} {\bibfnamefont {M.}~\bibnamefont {Singh}}, \bibinfo {author} {\bibfnamefont {H.~S.~J.}\ \bibnamefont {{van der Zant}}}, \bibinfo {author} {\bibfnamefont {P.~G.}\ \bibnamefont {Steeneken}}, \bibinfo {author} {\bibfnamefont {P.~M.}\ \bibnamefont {Sarro}},\ and\ \bibinfo {author} {\bibfnamefont {S.}~\bibnamefont {Vollebregt}},\ }\bibfield  {title} {\bibinfo {title} {Multi-layer graphene pirani pressure sensors},\ }\bibfield  {journal} {\bibinfo  {journal} {Nanotechnology}\ }\textbf {\bibinfo {volume} {32}},\ \href {https://doi.org/10.1088/1361-6528/abff8e} {10.1088/1361-6528/abff8e} (\bibinfo {year} {2021})\BibitemShut {NoStop}%
\bibitem [{\citenamefont {Shin}\ \emph {et~al.}(2023)\citenamefont {Shin}, \citenamefont {Kim}, \citenamefont {Kim}, \citenamefont {Cheong}, \citenamefont {Steeneken}, \citenamefont {Joo},\ and\ \citenamefont {Lee}}]{Shin.2023}%
  \BibitemOpen
  \bibfield  {author} {\bibinfo {author} {\bibfnamefont {D.~H.}\ \bibnamefont {Shin}}, \bibinfo {author} {\bibfnamefont {H.}~\bibnamefont {Kim}}, \bibinfo {author} {\bibfnamefont {S.~H.}\ \bibnamefont {Kim}}, \bibinfo {author} {\bibfnamefont {H.}~\bibnamefont {Cheong}}, \bibinfo {author} {\bibfnamefont {P.~G.}\ \bibnamefont {Steeneken}}, \bibinfo {author} {\bibfnamefont {C.}~\bibnamefont {Joo}},\ and\ \bibinfo {author} {\bibfnamefont {S.~W.}\ \bibnamefont {Lee}},\ }\bibfield  {title} {\bibinfo {title} {Graphene nano-electromechanical mass sensor with high resolution at room temperature},\ }\href {https://doi.org/10.1016/j.isci.2023.105958} {\bibfield  {journal} {\bibinfo  {journal} {iScience}\ }\textbf {\bibinfo {volume} {26}},\ \bibinfo {pages} {105958} (\bibinfo {year} {2023})}\BibitemShut {NoStop}%
\bibitem [{\citenamefont {Stoll}\ and\ \citenamefont {Kolmakov}(2012)}]{Stoll.2012}%
  \BibitemOpen
  \bibfield  {author} {\bibinfo {author} {\bibfnamefont {J.~D.}\ \bibnamefont {Stoll}}\ and\ \bibinfo {author} {\bibfnamefont {A.}~\bibnamefont {Kolmakov}},\ }\bibfield  {title} {\bibinfo {title} {Electron transparent graphene windows for environmental scanning electron microscopy in liquids and dense gases},\ }\href {https://doi.org/10.1088/0957-4484/23/50/505704} {\bibfield  {journal} {\bibinfo  {journal} {Nanotechnology}\ }\textbf {\bibinfo {volume} {23}},\ \bibinfo {pages} {505704} (\bibinfo {year} {2012})}\BibitemShut {NoStop}%
\bibitem [{\citenamefont {Leidinger}\ \emph {et~al.}(2021)\citenamefont {Leidinger}, \citenamefont {Kraus}, \citenamefont {Kratky}, \citenamefont {Zeller}, \citenamefont {Mente{\c{s}}}, \citenamefont {Genuzio}, \citenamefont {Locatelli},\ and\ \citenamefont {G{\"u}nther}}]{Leidinger.2021}%
  \BibitemOpen
  \bibfield  {author} {\bibinfo {author} {\bibfnamefont {P.}~\bibnamefont {Leidinger}}, \bibinfo {author} {\bibfnamefont {J.}~\bibnamefont {Kraus}}, \bibinfo {author} {\bibfnamefont {T.}~\bibnamefont {Kratky}}, \bibinfo {author} {\bibfnamefont {P.}~\bibnamefont {Zeller}}, \bibinfo {author} {\bibfnamefont {T.~O.}\ \bibnamefont {Mente{\c{s}}}}, \bibinfo {author} {\bibfnamefont {F.}~\bibnamefont {Genuzio}}, \bibinfo {author} {\bibfnamefont {A.}~\bibnamefont {Locatelli}},\ and\ \bibinfo {author} {\bibfnamefont {S.}~\bibnamefont {G{\"u}nther}},\ }\bibfield  {title} {\bibinfo {title} {Toward the perfect membrane material for environmental {X}-ray photoelectron spectroscopy},\ }\href {https://doi.org/10.1088/1361-6463/abe743} {\bibfield  {journal} {\bibinfo  {journal} {Journal of Physics D: Applied Physics}\ }\textbf {\bibinfo {volume} {54}},\ \bibinfo {pages} {234001} (\bibinfo {year} {2021})}\BibitemShut {NoStop}%
\bibitem [{\citenamefont {O'Hern}\ \emph {et~al.}(2012)\citenamefont {O'Hern}, \citenamefont {Stewart}, \citenamefont {Boutilier}, \citenamefont {Idrobo}, \citenamefont {Bhaviripudi}, \citenamefont {Das}, \citenamefont {Kong}, \citenamefont {Laoui}, \citenamefont {Atieh},\ and\ \citenamefont {Karnik}}]{OHern.2012}%
  \BibitemOpen
  \bibfield  {author} {\bibinfo {author} {\bibfnamefont {S.~C.}\ \bibnamefont {O'Hern}}, \bibinfo {author} {\bibfnamefont {C.~A.}\ \bibnamefont {Stewart}}, \bibinfo {author} {\bibfnamefont {M.~S.~H.}\ \bibnamefont {Boutilier}}, \bibinfo {author} {\bibfnamefont {J.-C.}\ \bibnamefont {Idrobo}}, \bibinfo {author} {\bibfnamefont {S.}~\bibnamefont {Bhaviripudi}}, \bibinfo {author} {\bibfnamefont {S.~K.}\ \bibnamefont {Das}}, \bibinfo {author} {\bibfnamefont {J.}~\bibnamefont {Kong}}, \bibinfo {author} {\bibfnamefont {T.}~\bibnamefont {Laoui}}, \bibinfo {author} {\bibfnamefont {M.}~\bibnamefont {Atieh}},\ and\ \bibinfo {author} {\bibfnamefont {R.}~\bibnamefont {Karnik}},\ }\bibfield  {title} {\bibinfo {title} {Selective molecular transport through intrinsic defects in a single layer of {CVD} graphene},\ }\href {https://doi.org/10.1021/nn303869m} {\bibfield  {journal} {\bibinfo  {journal} {ACS Nano}\ }\textbf {\bibinfo {volume} {6}},\ \bibinfo {pages} {10130} (\bibinfo {year} {2012})}\BibitemShut {NoStop}%
\bibitem [{\citenamefont {Koenig}\ \emph {et~al.}(2012)\citenamefont {Koenig}, \citenamefont {Wang}, \citenamefont {Pellegrino},\ and\ \citenamefont {Bunch}}]{Koenig.2012}%
  \BibitemOpen
  \bibfield  {author} {\bibinfo {author} {\bibfnamefont {S.~P.}\ \bibnamefont {Koenig}}, \bibinfo {author} {\bibfnamefont {L.}~\bibnamefont {Wang}}, \bibinfo {author} {\bibfnamefont {J.}~\bibnamefont {Pellegrino}},\ and\ \bibinfo {author} {\bibfnamefont {J.~S.}\ \bibnamefont {Bunch}},\ }\bibfield  {title} {\bibinfo {title} {Selective molecular sieving through porous graphene},\ }\href {https://doi.org/10.1038/nnano.2012.162} {\bibfield  {journal} {\bibinfo  {journal} {Nature Nanotechnology}\ }\textbf {\bibinfo {volume} {7}},\ \bibinfo {pages} {728} (\bibinfo {year} {2012})}\BibitemShut {NoStop}%
\bibitem [{\citenamefont {Surwade}\ \emph {et~al.}(2015)\citenamefont {Surwade}, \citenamefont {Smirnov}, \citenamefont {Vlassiouk}, \citenamefont {Unocic}, \citenamefont {Veith}, \citenamefont {Dai},\ and\ \citenamefont {Mahurin}}]{Surwade.2015}%
  \BibitemOpen
  \bibfield  {author} {\bibinfo {author} {\bibfnamefont {S.~P.}\ \bibnamefont {Surwade}}, \bibinfo {author} {\bibfnamefont {S.~N.}\ \bibnamefont {Smirnov}}, \bibinfo {author} {\bibfnamefont {I.~V.}\ \bibnamefont {Vlassiouk}}, \bibinfo {author} {\bibfnamefont {R.~R.}\ \bibnamefont {Unocic}}, \bibinfo {author} {\bibfnamefont {G.~M.}\ \bibnamefont {Veith}}, \bibinfo {author} {\bibfnamefont {S.}~\bibnamefont {Dai}},\ and\ \bibinfo {author} {\bibfnamefont {S.~M.}\ \bibnamefont {Mahurin}},\ }\bibfield  {title} {\bibinfo {title} {Water desalination using nanoporous single-layer graphene},\ }\href {https://doi.org/10.1038/nnano.2015.37} {\bibfield  {journal} {\bibinfo  {journal} {Nature Nanotechnology}\ }\textbf {\bibinfo {volume} {10}},\ \bibinfo {pages} {459} (\bibinfo {year} {2015})}\BibitemShut {NoStop}%
\bibitem [{\citenamefont {Madau{\ss}}\ \emph {et~al.}(2017)\citenamefont {Madau{\ss}}, \citenamefont {Schumacher}, \citenamefont {Ghosh}, \citenamefont {Ochedowski}, \citenamefont {Meyer}, \citenamefont {Lebius}, \citenamefont {Ban-d'Etat}, \citenamefont {Toimil-Molares}, \citenamefont {Trautmann}, \citenamefont {Lammertink}, \citenamefont {Ulbricht},\ and\ \citenamefont {Schleberger}}]{Madau.2017}%
  \BibitemOpen
  \bibfield  {author} {\bibinfo {author} {\bibfnamefont {L.}~\bibnamefont {Madau{\ss}}}, \bibinfo {author} {\bibfnamefont {J.}~\bibnamefont {Schumacher}}, \bibinfo {author} {\bibfnamefont {M.}~\bibnamefont {Ghosh}}, \bibinfo {author} {\bibfnamefont {O.}~\bibnamefont {Ochedowski}}, \bibinfo {author} {\bibfnamefont {J.}~\bibnamefont {Meyer}}, \bibinfo {author} {\bibfnamefont {H.}~\bibnamefont {Lebius}}, \bibinfo {author} {\bibfnamefont {B.}~\bibnamefont {Ban-d'Etat}}, \bibinfo {author} {\bibfnamefont {M.~E.}\ \bibnamefont {Toimil-Molares}}, \bibinfo {author} {\bibfnamefont {C.}~\bibnamefont {Trautmann}}, \bibinfo {author} {\bibfnamefont {R.~G.~H.}\ \bibnamefont {Lammertink}}, \bibinfo {author} {\bibfnamefont {M.}~\bibnamefont {Ulbricht}},\ and\ \bibinfo {author} {\bibfnamefont {M.}~\bibnamefont {Schleberger}},\ }\bibfield  {title} {\bibinfo {title} {Fabrication of nanoporous graphene/polymer composite membranes},\ }\href {https://doi.org/10.1039/C7NR02755A} {\bibfield  {journal} {\bibinfo  {journal}
  {Nanoscale}\ }\textbf {\bibinfo {volume} {9}},\ \bibinfo {pages} {10487} (\bibinfo {year} {2017})}\BibitemShut {NoStop}%
\bibitem [{\citenamefont {Frank}\ \emph {et~al.}(2007)\citenamefont {Frank}, \citenamefont {Tanenbaum}, \citenamefont {{van der Zande}},\ and\ \citenamefont {McEuen}}]{Frank.2007}%
  \BibitemOpen
  \bibfield  {author} {\bibinfo {author} {\bibfnamefont {I.~W.}\ \bibnamefont {Frank}}, \bibinfo {author} {\bibfnamefont {D.~M.}\ \bibnamefont {Tanenbaum}}, \bibinfo {author} {\bibfnamefont {A.~M.}\ \bibnamefont {{van der Zande}}},\ and\ \bibinfo {author} {\bibfnamefont {P.~L.}\ \bibnamefont {McEuen}},\ }\bibfield  {title} {\bibinfo {title} {Mechanical properties of suspended graphene sheets},\ }\href {https://doi.org/10.1116/1.2789446} {\bibfield  {journal} {\bibinfo  {journal} {Journal of Vacuum Science {\&} Technology B: Microelectronics and Nanometer Structures Processing, Measurement, and Phenomena}\ }\textbf {\bibinfo {volume} {25}},\ \bibinfo {pages} {2558} (\bibinfo {year} {2007})}\BibitemShut {NoStop}%
\bibitem [{\citenamefont {Lee}\ \emph {et~al.}(2008)\citenamefont {Lee}, \citenamefont {Wei}, \citenamefont {Kysar},\ and\ \citenamefont {Hone}}]{Lee.2008}%
  \BibitemOpen
  \bibfield  {author} {\bibinfo {author} {\bibfnamefont {C.}~\bibnamefont {Lee}}, \bibinfo {author} {\bibfnamefont {X.}~\bibnamefont {Wei}}, \bibinfo {author} {\bibfnamefont {J.~W.}\ \bibnamefont {Kysar}},\ and\ \bibinfo {author} {\bibfnamefont {J.}~\bibnamefont {Hone}},\ }\bibfield  {title} {\bibinfo {title} {Measurement of the elastic properties and intrinsic strength of monolayer graphene},\ }\href {https://doi.org/10.1126/science.1157996} {\bibfield  {journal} {\bibinfo  {journal} {Science (New York, N.Y.)}\ }\textbf {\bibinfo {volume} {321}},\ \bibinfo {pages} {385} (\bibinfo {year} {2008})}\BibitemShut {NoStop}%
\bibitem [{\citenamefont {Booth}\ \emph {et~al.}(2008)\citenamefont {Booth}, \citenamefont {Blake}, \citenamefont {Nair}, \citenamefont {{Da Jiang}}, \citenamefont {Hill}, \citenamefont {Bangert}, \citenamefont {Bleloch}, \citenamefont {Gass}, \citenamefont {Novoselov}, \citenamefont {Katsnelson},\ and\ \citenamefont {Geim}}]{Booth.2008}%
  \BibitemOpen
  \bibfield  {author} {\bibinfo {author} {\bibfnamefont {T.~J.}\ \bibnamefont {Booth}}, \bibinfo {author} {\bibfnamefont {P.}~\bibnamefont {Blake}}, \bibinfo {author} {\bibfnamefont {R.~R.}\ \bibnamefont {Nair}}, \bibinfo {author} {\bibnamefont {{Da Jiang}}}, \bibinfo {author} {\bibfnamefont {E.~W.}\ \bibnamefont {Hill}}, \bibinfo {author} {\bibfnamefont {U.}~\bibnamefont {Bangert}}, \bibinfo {author} {\bibfnamefont {A.}~\bibnamefont {Bleloch}}, \bibinfo {author} {\bibfnamefont {M.}~\bibnamefont {Gass}}, \bibinfo {author} {\bibfnamefont {K.~S.}\ \bibnamefont {Novoselov}}, \bibinfo {author} {\bibfnamefont {M.~I.}\ \bibnamefont {Katsnelson}},\ and\ \bibinfo {author} {\bibfnamefont {A.~K.}\ \bibnamefont {Geim}},\ }\bibfield  {title} {\bibinfo {title} {Macroscopic graphene membranes and their extraordinary stiffness},\ }\href {https://doi.org/10.1021/nl801412y} {\bibfield  {journal} {\bibinfo  {journal} {Nano letters}\ }\textbf {\bibinfo {volume} {8}},\ \bibinfo {pages} {2442} (\bibinfo {year}
  {2008})}\BibitemShut {NoStop}%
\bibitem [{\citenamefont {Ochedowski}\ \emph {et~al.}(2013)\citenamefont {Ochedowski}, \citenamefont {Marinov}, \citenamefont {Wilbs}, \citenamefont {Keller}, \citenamefont {Scheuschner}, \citenamefont {Severin}, \citenamefont {Bender}, \citenamefont {Maultzsch}, \citenamefont {Tegude},\ and\ \citenamefont {Schleberger}}]{Ochedowski.2013}%
  \BibitemOpen
  \bibfield  {author} {\bibinfo {author} {\bibfnamefont {O.}~\bibnamefont {Ochedowski}}, \bibinfo {author} {\bibfnamefont {K.}~\bibnamefont {Marinov}}, \bibinfo {author} {\bibfnamefont {G.}~\bibnamefont {Wilbs}}, \bibinfo {author} {\bibfnamefont {G.}~\bibnamefont {Keller}}, \bibinfo {author} {\bibfnamefont {N.}~\bibnamefont {Scheuschner}}, \bibinfo {author} {\bibfnamefont {D.}~\bibnamefont {Severin}}, \bibinfo {author} {\bibfnamefont {M.}~\bibnamefont {Bender}}, \bibinfo {author} {\bibfnamefont {J.}~\bibnamefont {Maultzsch}}, \bibinfo {author} {\bibfnamefont {F.~J.}\ \bibnamefont {Tegude}},\ and\ \bibinfo {author} {\bibfnamefont {M.}~\bibnamefont {Schleberger}},\ }\bibfield  {title} {\bibinfo {title} {Radiation hardness of graphene and {MoS}$_2$ field effect devices against swift heavy ion irradiation},\ }\bibfield  {journal} {\bibinfo  {journal} {Journal of Applied Physics}\ }\textbf {\bibinfo {volume} {113}},\ \href {https://doi.org/10.1063/1.4808460} {10.1063/1.4808460} (\bibinfo {year} {2013})\BibitemShut
  {NoStop}%
\bibitem [{\citenamefont {Kozubek}\ \emph {et~al.}(2014)\citenamefont {Kozubek}, \citenamefont {Ochedowski}, \citenamefont {Zagoranskiy}, \citenamefont {Karlu{\v{s}}i{\'c}},\ and\ \citenamefont {Schleberger}}]{Kozubek.2014}%
  \BibitemOpen
  \bibfield  {author} {\bibinfo {author} {\bibfnamefont {R.}~\bibnamefont {Kozubek}}, \bibinfo {author} {\bibfnamefont {O.}~\bibnamefont {Ochedowski}}, \bibinfo {author} {\bibfnamefont {I.}~\bibnamefont {Zagoranskiy}}, \bibinfo {author} {\bibfnamefont {M.}~\bibnamefont {Karlu{\v{s}}i{\'c}}},\ and\ \bibinfo {author} {\bibfnamefont {M.}~\bibnamefont {Schleberger}},\ }\bibfield  {title} {\bibinfo {title} {Application of {HOPG} and {CVD} graphene as ion beam detectors},\ }\href {https://doi.org/10.1016/j.nimb.2014.07.034} {\bibfield  {journal} {\bibinfo  {journal} {Nuclear Instruments and Methods in Physics Research Section B: Beam Interactions with Materials and Atoms}\ }\textbf {\bibinfo {volume} {340}},\ \bibinfo {pages} {81} (\bibinfo {year} {2014})}\BibitemShut {NoStop}%
\bibitem [{\citenamefont {Ernst}\ \emph {et~al.}(2016)\citenamefont {Ernst}, \citenamefont {Kozubek}, \citenamefont {Madau{\ss}}, \citenamefont {Sonntag}, \citenamefont {Lorke},\ and\ \citenamefont {Schleberger}}]{Ernst.2016}%
  \BibitemOpen
  \bibfield  {author} {\bibinfo {author} {\bibfnamefont {P.}~\bibnamefont {Ernst}}, \bibinfo {author} {\bibfnamefont {R.}~\bibnamefont {Kozubek}}, \bibinfo {author} {\bibfnamefont {L.}~\bibnamefont {Madau{\ss}}}, \bibinfo {author} {\bibfnamefont {J.}~\bibnamefont {Sonntag}}, \bibinfo {author} {\bibfnamefont {A.}~\bibnamefont {Lorke}},\ and\ \bibinfo {author} {\bibfnamefont {M.}~\bibnamefont {Schleberger}},\ }\bibfield  {title} {\bibinfo {title} {Irradiation of graphene field effect transistors with highly charged ions},\ }\href {https://doi.org/10.1016/j.nimb.2016.03.043} {\bibfield  {journal} {\bibinfo  {journal} {Nuclear Instruments and Methods in Physics Research Section B: Beam Interactions with Materials and Atoms}\ }\textbf {\bibinfo {volume} {382}},\ \bibinfo {pages} {71} (\bibinfo {year} {2016})}\BibitemShut {NoStop}%
\bibitem [{\citenamefont {Gruber}\ \emph {et~al.}(2016)\citenamefont {Gruber}, \citenamefont {Wilhelm}, \citenamefont {P{\'e}tuya}, \citenamefont {Smejkal}, \citenamefont {Kozubek}, \citenamefont {Hierzenberger}, \citenamefont {Bayer}, \citenamefont {Aldazabal}, \citenamefont {Kazansky}, \citenamefont {Libisch}, \citenamefont {Krasheninnikov}, \citenamefont {Schleberger}, \citenamefont {Facsko}, \citenamefont {Borisov}, \citenamefont {Arnau},\ and\ \citenamefont {Aumayr}}]{Gruber.2016}%
  \BibitemOpen
  \bibfield  {author} {\bibinfo {author} {\bibfnamefont {E.}~\bibnamefont {Gruber}}, \bibinfo {author} {\bibfnamefont {R.~A.}\ \bibnamefont {Wilhelm}}, \bibinfo {author} {\bibfnamefont {R.}~\bibnamefont {P{\'e}tuya}}, \bibinfo {author} {\bibfnamefont {V.}~\bibnamefont {Smejkal}}, \bibinfo {author} {\bibfnamefont {R.}~\bibnamefont {Kozubek}}, \bibinfo {author} {\bibfnamefont {A.}~\bibnamefont {Hierzenberger}}, \bibinfo {author} {\bibfnamefont {B.~C.}\ \bibnamefont {Bayer}}, \bibinfo {author} {\bibfnamefont {I.}~\bibnamefont {Aldazabal}}, \bibinfo {author} {\bibfnamefont {A.~K.}\ \bibnamefont {Kazansky}}, \bibinfo {author} {\bibfnamefont {F.}~\bibnamefont {Libisch}}, \bibinfo {author} {\bibfnamefont {A.~V.}\ \bibnamefont {Krasheninnikov}}, \bibinfo {author} {\bibfnamefont {M.}~\bibnamefont {Schleberger}}, \bibinfo {author} {\bibfnamefont {S.}~\bibnamefont {Facsko}}, \bibinfo {author} {\bibfnamefont {A.~G.}\ \bibnamefont {Borisov}}, \bibinfo {author} {\bibfnamefont {A.}~\bibnamefont {Arnau}},\ and\ \bibinfo
  {author} {\bibfnamefont {F.}~\bibnamefont {Aumayr}},\ }\bibfield  {title} {\bibinfo {title} {Ultrafast electronic response of graphene to a strong and localized electric field},\ }\href {https://doi.org/10.1038/ncomms13948} {\bibfield  {journal} {\bibinfo  {journal} {Nature Communications}\ }\textbf {\bibinfo {volume} {7}},\ \bibinfo {pages} {13948} (\bibinfo {year} {2016})}\BibitemShut {NoStop}%
\bibitem [{\citenamefont {Kuramitsu}\ \emph {et~al.}(2022)\citenamefont {Kuramitsu}, \citenamefont {Minami}, \citenamefont {Hihara}, \citenamefont {Sakai}, \citenamefont {Nishimoto}, \citenamefont {Isayama}, \citenamefont {Liao}, \citenamefont {Wu}, \citenamefont {Woon}, \citenamefont {Chen}, \citenamefont {Liu}, \citenamefont {He}, \citenamefont {Su}, \citenamefont {Ota}, \citenamefont {Egashira}, \citenamefont {Morace}, \citenamefont {Sakawa}, \citenamefont {Abe}, \citenamefont {Habara}, \citenamefont {Kodama}, \citenamefont {D{\"o}hl}, \citenamefont {Woolsey}, \citenamefont {Koenig}, \citenamefont {Kumar}, \citenamefont {Ohnishi}, \citenamefont {Kanasaki}, \citenamefont {Asai}, \citenamefont {Yamauchi}, \citenamefont {Oda}, \citenamefont {Kondo}, \citenamefont {Kiriyama},\ and\ \citenamefont {Fukuda}}]{Kuramitsu.2022}%
  \BibitemOpen
  \bibfield  {author} {\bibinfo {author} {\bibfnamefont {Y.}~\bibnamefont {Kuramitsu}}, \bibinfo {author} {\bibfnamefont {T.}~\bibnamefont {Minami}}, \bibinfo {author} {\bibfnamefont {T.}~\bibnamefont {Hihara}}, \bibinfo {author} {\bibfnamefont {K.}~\bibnamefont {Sakai}}, \bibinfo {author} {\bibfnamefont {T.}~\bibnamefont {Nishimoto}}, \bibinfo {author} {\bibfnamefont {S.}~\bibnamefont {Isayama}}, \bibinfo {author} {\bibfnamefont {Y.~T.}\ \bibnamefont {Liao}}, \bibinfo {author} {\bibfnamefont {K.~T.}\ \bibnamefont {Wu}}, \bibinfo {author} {\bibfnamefont {W.~Y.}\ \bibnamefont {Woon}}, \bibinfo {author} {\bibfnamefont {S.~H.}\ \bibnamefont {Chen}}, \bibinfo {author} {\bibfnamefont {Y.~L.}\ \bibnamefont {Liu}}, \bibinfo {author} {\bibfnamefont {S.~M.}\ \bibnamefont {He}}, \bibinfo {author} {\bibfnamefont {C.~Y.}\ \bibnamefont {Su}}, \bibinfo {author} {\bibfnamefont {M.}~\bibnamefont {Ota}}, \bibinfo {author} {\bibfnamefont {S.}~\bibnamefont {Egashira}}, \bibinfo {author} {\bibfnamefont {A.}~\bibnamefont
  {Morace}}, \bibinfo {author} {\bibfnamefont {Y.}~\bibnamefont {Sakawa}}, \bibinfo {author} {\bibfnamefont {Y.}~\bibnamefont {Abe}}, \bibinfo {author} {\bibfnamefont {H.}~\bibnamefont {Habara}}, \bibinfo {author} {\bibfnamefont {R.}~\bibnamefont {Kodama}}, \bibinfo {author} {\bibfnamefont {L.~N.~K.}\ \bibnamefont {D{\"o}hl}}, \bibinfo {author} {\bibfnamefont {N.}~\bibnamefont {Woolsey}}, \bibinfo {author} {\bibfnamefont {M.}~\bibnamefont {Koenig}}, \bibinfo {author} {\bibfnamefont {H.~S.}\ \bibnamefont {Kumar}}, \bibinfo {author} {\bibfnamefont {N.}~\bibnamefont {Ohnishi}}, \bibinfo {author} {\bibfnamefont {M.}~\bibnamefont {Kanasaki}}, \bibinfo {author} {\bibfnamefont {T.}~\bibnamefont {Asai}}, \bibinfo {author} {\bibfnamefont {T.}~\bibnamefont {Yamauchi}}, \bibinfo {author} {\bibfnamefont {K.}~\bibnamefont {Oda}}, \bibinfo {author} {\bibfnamefont {K.}~\bibnamefont {Kondo}}, \bibinfo {author} {\bibfnamefont {H.}~\bibnamefont {Kiriyama}},\ and\ \bibinfo {author} {\bibfnamefont {Y.}~\bibnamefont {Fukuda}},\
  }\bibfield  {title} {\bibinfo {title} {Robustness of large-area suspended graphene under interaction with intense laser},\ }\href {https://doi.org/10.1038/s41598-022-06055-4} {\bibfield  {journal} {\bibinfo  {journal} {Scientific Reports}\ }\textbf {\bibinfo {volume} {12}},\ \bibinfo {pages} {2346} (\bibinfo {year} {2022})}\BibitemShut {NoStop}%
\bibitem [{\citenamefont {Zan}\ \emph {et~al.}(2012)\citenamefont {Zan}, \citenamefont {Ramasse}, \citenamefont {Bangert},\ and\ \citenamefont {Novoselov}}]{Zan.2012}%
  \BibitemOpen
  \bibfield  {author} {\bibinfo {author} {\bibfnamefont {R.}~\bibnamefont {Zan}}, \bibinfo {author} {\bibfnamefont {Q.~M.}\ \bibnamefont {Ramasse}}, \bibinfo {author} {\bibfnamefont {U.}~\bibnamefont {Bangert}},\ and\ \bibinfo {author} {\bibfnamefont {K.~S.}\ \bibnamefont {Novoselov}},\ }\bibfield  {title} {\bibinfo {title} {Graphene reknits its holes},\ }\href {https://doi.org/10.1021/nl300985q} {\bibfield  {journal} {\bibinfo  {journal} {Nano letters}\ }\textbf {\bibinfo {volume} {12}},\ \bibinfo {pages} {3936} (\bibinfo {year} {2012})}\BibitemShut {NoStop}%
\bibitem [{\citenamefont {Creutzburg}\ \emph {et~al.}(2021)\citenamefont {Creutzburg}, \citenamefont {Niggas}, \citenamefont {Weichselbaum}, \citenamefont {Grande}, \citenamefont {Aumayr},\ and\ \citenamefont {Wilhelm}}]{Creutzburg.2021}%
  \BibitemOpen
  \bibfield  {author} {\bibinfo {author} {\bibfnamefont {S.}~\bibnamefont {Creutzburg}}, \bibinfo {author} {\bibfnamefont {A.}~\bibnamefont {Niggas}}, \bibinfo {author} {\bibfnamefont {D.}~\bibnamefont {Weichselbaum}}, \bibinfo {author} {\bibfnamefont {P.~L.}\ \bibnamefont {Grande}}, \bibinfo {author} {\bibfnamefont {F.}~\bibnamefont {Aumayr}},\ and\ \bibinfo {author} {\bibfnamefont {R.~A.}\ \bibnamefont {Wilhelm}},\ }\bibfield  {title} {\bibinfo {title} {Angle-dependent charge exchange and energy loss of slow highly charged ions in freestanding graphene},\ }\bibfield  {journal} {\bibinfo  {journal} {Physical Review A}\ }\textbf {\bibinfo {volume} {104}},\ \href {https://doi.org/10.1103/PhysRevA.104.042806} {10.1103/PhysRevA.104.042806} (\bibinfo {year} {2021})\BibitemShut {NoStop}%
\bibitem [{\citenamefont {Tilmann}\ \emph {et~al.}(2023)\citenamefont {Tilmann}, \citenamefont {Bartlam}, \citenamefont {Hartwig}, \citenamefont {Tywoniuk}, \citenamefont {Dominik}, \citenamefont {Cullen}, \citenamefont {Peters}, \citenamefont {Stimpel-Lindner}, \citenamefont {McEvoy},\ and\ \citenamefont {Duesberg}}]{Tilmann.2023}%
  \BibitemOpen
  \bibfield  {author} {\bibinfo {author} {\bibfnamefont {R.}~\bibnamefont {Tilmann}}, \bibinfo {author} {\bibfnamefont {C.}~\bibnamefont {Bartlam}}, \bibinfo {author} {\bibfnamefont {O.}~\bibnamefont {Hartwig}}, \bibinfo {author} {\bibfnamefont {B.}~\bibnamefont {Tywoniuk}}, \bibinfo {author} {\bibfnamefont {N.}~\bibnamefont {Dominik}}, \bibinfo {author} {\bibfnamefont {C.~P.}\ \bibnamefont {Cullen}}, \bibinfo {author} {\bibfnamefont {L.}~\bibnamefont {Peters}}, \bibinfo {author} {\bibfnamefont {T.}~\bibnamefont {Stimpel-Lindner}}, \bibinfo {author} {\bibfnamefont {N.}~\bibnamefont {McEvoy}},\ and\ \bibinfo {author} {\bibfnamefont {G.~S.}\ \bibnamefont {Duesberg}},\ }\bibfield  {title} {\bibinfo {title} {Identification of ubiquitously present polymeric adlayers on {2D} transition metal dichalcogenides},\ }\href {https://doi.org/10.1021/acsnano.3c01649} {\bibfield  {journal} {\bibinfo  {journal} {ACS Nano}\ }\textbf {\bibinfo {volume} {17}},\ \bibinfo {pages} {10617} (\bibinfo {year} {2023})}\BibitemShut
  {NoStop}%
\bibitem [{\citenamefont {Madau{\ss}}\ \emph {et~al.}(2020)\citenamefont {Madau{\ss}}, \citenamefont {Pollmann}, \citenamefont {Foller}, \citenamefont {Schumacher}, \citenamefont {Hagemann}, \citenamefont {Heckhoff}, \citenamefont {Herder}, \citenamefont {Skopinski}, \citenamefont {Breuer}, \citenamefont {Hierzenberger}, \citenamefont {Wittmar}, \citenamefont {Lebius}, \citenamefont {Benyagoub}, \citenamefont {Ulbricht}, \citenamefont {Joshi},\ and\ \citenamefont {Schleberger}}]{Madau.2020}%
  \BibitemOpen
  \bibfield  {author} {\bibinfo {author} {\bibfnamefont {L.}~\bibnamefont {Madau{\ss}}}, \bibinfo {author} {\bibfnamefont {E.}~\bibnamefont {Pollmann}}, \bibinfo {author} {\bibfnamefont {T.}~\bibnamefont {Foller}}, \bibinfo {author} {\bibfnamefont {J.}~\bibnamefont {Schumacher}}, \bibinfo {author} {\bibfnamefont {U.}~\bibnamefont {Hagemann}}, \bibinfo {author} {\bibfnamefont {T.}~\bibnamefont {Heckhoff}}, \bibinfo {author} {\bibfnamefont {M.}~\bibnamefont {Herder}}, \bibinfo {author} {\bibfnamefont {L.}~\bibnamefont {Skopinski}}, \bibinfo {author} {\bibfnamefont {L.}~\bibnamefont {Breuer}}, \bibinfo {author} {\bibfnamefont {A.}~\bibnamefont {Hierzenberger}}, \bibinfo {author} {\bibfnamefont {A.}~\bibnamefont {Wittmar}}, \bibinfo {author} {\bibfnamefont {H.}~\bibnamefont {Lebius}}, \bibinfo {author} {\bibfnamefont {A.}~\bibnamefont {Benyagoub}}, \bibinfo {author} {\bibfnamefont {M.}~\bibnamefont {Ulbricht}}, \bibinfo {author} {\bibfnamefont {R.}~\bibnamefont {Joshi}},\ and\ \bibinfo {author} {\bibfnamefont
  {M.}~\bibnamefont {Schleberger}},\ }\bibfield  {title} {\bibinfo {title} {A swift technique to hydrophobize graphene and increase its mechanical stability and charge carrier density},\ }\href {https://doi.org/10.1038/s41699-020-0148-9} {\bibfield  {journal} {\bibinfo  {journal} {npj 2D Materials and Applications}\ }\textbf {\bibinfo {volume} {4}},\ \bibinfo {pages} {1} (\bibinfo {year} {2020})}\BibitemShut {NoStop}%
\bibitem [{\citenamefont {Huang}\ \emph {et~al.}(2011)\citenamefont {Huang}, \citenamefont {Ruiz-Vargas}, \citenamefont {{van der Zande}}, \citenamefont {Whitney}, \citenamefont {Levendorf}, \citenamefont {ke{V}ek}, \citenamefont {Garg}, \citenamefont {Alden}, \citenamefont {Hustedt}, \citenamefont {Zhu}, \citenamefont {Park}, \citenamefont {McEuen},\ and\ \citenamefont {Muller}}]{Huang.2011}%
  \BibitemOpen
  \bibfield  {author} {\bibinfo {author} {\bibfnamefont {P.~Y.}\ \bibnamefont {Huang}}, \bibinfo {author} {\bibfnamefont {C.~S.}\ \bibnamefont {Ruiz-Vargas}}, \bibinfo {author} {\bibfnamefont {A.~M.}\ \bibnamefont {{van der Zande}}}, \bibinfo {author} {\bibfnamefont {W.~S.}\ \bibnamefont {Whitney}}, \bibinfo {author} {\bibfnamefont {M.~P.}\ \bibnamefont {Levendorf}}, \bibinfo {author} {\bibfnamefont {J.~W.}\ \bibnamefont {ke{V}ek}}, \bibinfo {author} {\bibfnamefont {S.}~\bibnamefont {Garg}}, \bibinfo {author} {\bibfnamefont {J.~S.}\ \bibnamefont {Alden}}, \bibinfo {author} {\bibfnamefont {C.~J.}\ \bibnamefont {Hustedt}}, \bibinfo {author} {\bibfnamefont {Y.}~\bibnamefont {Zhu}}, \bibinfo {author} {\bibfnamefont {J.}~\bibnamefont {Park}}, \bibinfo {author} {\bibfnamefont {P.~L.}\ \bibnamefont {McEuen}},\ and\ \bibinfo {author} {\bibfnamefont {D.~A.}\ \bibnamefont {Muller}},\ }\bibfield  {title} {\bibinfo {title} {Grains and grain boundaries in single-layer graphene atomic patchwork quilts},\ }\href
  {https://doi.org/10.1038/nature09718} {\bibfield  {journal} {\bibinfo  {journal} {Nature}\ }\textbf {\bibinfo {volume} {469}},\ \bibinfo {pages} {389} (\bibinfo {year} {2011})}\BibitemShut {NoStop}%
\bibitem [{\citenamefont {Baragiola}\ \emph {et~al.}(1998)\citenamefont {Baragiola}, \citenamefont {Shi}, \citenamefont {Vidal},\ and\ \citenamefont {Dukes}}]{Baragiola.1998}%
  \BibitemOpen
  \bibfield  {author} {\bibinfo {author} {\bibfnamefont {R.~A.}\ \bibnamefont {Baragiola}}, \bibinfo {author} {\bibfnamefont {M.}~\bibnamefont {Shi}}, \bibinfo {author} {\bibfnamefont {R.~A.}\ \bibnamefont {Vidal}},\ and\ \bibinfo {author} {\bibfnamefont {C.~A.}\ \bibnamefont {Dukes}},\ }\bibfield  {title} {\bibinfo {title} {Fast proton-induced electron emission from rare-gas solids and electrostatic charging effects},\ }\href {https://doi.org/10.1103/PhysRevB.58.13212} {\bibfield  {journal} {\bibinfo  {journal} {Physical Review B}\ }\textbf {\bibinfo {volume} {58}},\ \bibinfo {pages} {13212} (\bibinfo {year} {1998})}\BibitemShut {NoStop}%
\bibitem [{\citenamefont {Lee}\ \emph {et~al.}(2012)\citenamefont {Lee}, \citenamefont {Ahn}, \citenamefont {Shim}, \citenamefont {Lee},\ and\ \citenamefont {Ryu}}]{Lee.2012}%
  \BibitemOpen
  \bibfield  {author} {\bibinfo {author} {\bibfnamefont {J.~E.}\ \bibnamefont {Lee}}, \bibinfo {author} {\bibfnamefont {G.}~\bibnamefont {Ahn}}, \bibinfo {author} {\bibfnamefont {J.}~\bibnamefont {Shim}}, \bibinfo {author} {\bibfnamefont {Y.~S.}\ \bibnamefont {Lee}},\ and\ \bibinfo {author} {\bibfnamefont {S.}~\bibnamefont {Ryu}},\ }\bibfield  {title} {\bibinfo {title} {Optical separation of mechanical strain from charge doping in graphene},\ }\href {https://doi.org/10.1038/ncomms2022} {\bibfield  {journal} {\bibinfo  {journal} {Nature Communications}\ }\textbf {\bibinfo {volume} {3}},\ \bibinfo {pages} {1024} (\bibinfo {year} {2012})}\BibitemShut {NoStop}%
\bibitem [{\citenamefont {Kim}\ and\ \citenamefont {Ryu}(2016)}]{Kim.2016}%
  \BibitemOpen
  \bibfield  {author} {\bibinfo {author} {\bibfnamefont {S.}~\bibnamefont {Kim}}\ and\ \bibinfo {author} {\bibfnamefont {S.}~\bibnamefont {Ryu}},\ }\bibfield  {title} {\bibinfo {title} {Thickness-dependent native strain in graphene membranes visualized by raman spectroscopy},\ }\href {https://doi.org/10.1016/j.carbon.2016.01.001} {\bibfield  {journal} {\bibinfo  {journal} {Carbon}\ }\textbf {\bibinfo {volume} {100}},\ \bibinfo {pages} {283} (\bibinfo {year} {2016})}\BibitemShut {NoStop}%
\bibitem [{\citenamefont {Ne{\v{c}}as}\ and\ \citenamefont {Klapetek}(2012)}]{Necas.2012}%
  \BibitemOpen
  \bibfield  {author} {\bibinfo {author} {\bibfnamefont {D.}~\bibnamefont {Ne{\v{c}}as}}\ and\ \bibinfo {author} {\bibfnamefont {P.}~\bibnamefont {Klapetek}},\ }\bibfield  {title} {\bibinfo {title} {Gwyddion: an open-source software for spm data analysis},\ }\bibfield  {journal} {\bibinfo  {journal} {Open Physics}\ }\textbf {\bibinfo {volume} {10}},\ \href {https://doi.org/10.2478/s11534-011-0096-2} {10.2478/s11534-011-0096-2} (\bibinfo {year} {2012})\BibitemShut {NoStop}%
\bibitem [{\citenamefont {Lloyd-Hughes}\ and\ \citenamefont {Jeon}(2012)}]{LloydHughes.2012}%
  \BibitemOpen
  \bibfield  {author} {\bibinfo {author} {\bibfnamefont {J.}~\bibnamefont {Lloyd-Hughes}}\ and\ \bibinfo {author} {\bibfnamefont {T.-I.}\ \bibnamefont {Jeon}},\ }\bibfield  {title} {\bibinfo {title} {A review of the terahertz conductivity of bulk and nano-materials},\ }\href {https://doi.org/10.1007/s10762-012-9905-y} {\bibfield  {journal} {\bibinfo  {journal} {Journal of Infrared, Millimeter, and Terahertz Waves}\ }\textbf {\bibinfo {volume} {33}},\ \bibinfo {pages} {871} (\bibinfo {year} {2012})}\BibitemShut {NoStop}%
\bibitem [{\citenamefont {Dawlaty}\ \emph {et~al.}(2008)\citenamefont {Dawlaty}, \citenamefont {Shivaraman}, \citenamefont {Strait}, \citenamefont {George}, \citenamefont {Chandrashekhar}, \citenamefont {Rana}, \citenamefont {Spencer}, \citenamefont {Veksler},\ and\ \citenamefont {Chen}}]{Dawlaty.2008}%
  \BibitemOpen
  \bibfield  {author} {\bibinfo {author} {\bibfnamefont {J.~M.}\ \bibnamefont {Dawlaty}}, \bibinfo {author} {\bibfnamefont {S.}~\bibnamefont {Shivaraman}}, \bibinfo {author} {\bibfnamefont {J.}~\bibnamefont {Strait}}, \bibinfo {author} {\bibfnamefont {P.}~\bibnamefont {George}}, \bibinfo {author} {\bibfnamefont {M.}~\bibnamefont {Chandrashekhar}}, \bibinfo {author} {\bibfnamefont {F.}~\bibnamefont {Rana}}, \bibinfo {author} {\bibfnamefont {M.~G.}\ \bibnamefont {Spencer}}, \bibinfo {author} {\bibfnamefont {D.}~\bibnamefont {Veksler}},\ and\ \bibinfo {author} {\bibfnamefont {Y.}~\bibnamefont {Chen}},\ }\bibfield  {title} {\bibinfo {title} {Measurement of the optical absorption spectra of epitaxial graphene from terahertz to visible},\ }\bibfield  {journal} {\bibinfo  {journal} {Applied Physics Letters}\ }\textbf {\bibinfo {volume} {93}},\ \href {https://doi.org/10.1063/1.2990753} {10.1063/1.2990753} (\bibinfo {year} {2008})\BibitemShut {NoStop}%
\bibitem [{\citenamefont {Adam}(2011)}]{Adam.2011}%
  \BibitemOpen
  \bibfield  {author} {\bibinfo {author} {\bibfnamefont {A.~J.~L.}\ \bibnamefont {Adam}},\ }\bibfield  {title} {\bibinfo {title} {Review of near-field terahertz measurement methods and their applications},\ }\href {https://doi.org/10.1007/s10762-011-9809-2} {\bibfield  {journal} {\bibinfo  {journal} {Journal of Infrared, Millimeter, and Terahertz Waves}\ }\textbf {\bibinfo {volume} {32}},\ \bibinfo {pages} {976} (\bibinfo {year} {2011})}\BibitemShut {NoStop}%
\bibitem [{\citenamefont {Glover}\ and\ \citenamefont {Tinkham}(1957)}]{Glover.1957}%
  \BibitemOpen
  \bibfield  {author} {\bibinfo {author} {\bibfnamefont {R.~E.}\ \bibnamefont {Glover}}\ and\ \bibinfo {author} {\bibfnamefont {M.}~\bibnamefont {Tinkham}},\ }\bibfield  {title} {\bibinfo {title} {Conductivity of superconducting films for photon energies between 0.3 and 40{kTc}},\ }\href {https://doi.org/10.1103/PhysRev.108.243} {\bibfield  {journal} {\bibinfo  {journal} {Physical Review}\ }\textbf {\bibinfo {volume} {108}},\ \bibinfo {pages} {243} (\bibinfo {year} {1957})}\BibitemShut {NoStop}%
\bibitem [{\citenamefont {Bolotin}\ \emph {et~al.}(2008)\citenamefont {Bolotin}, \citenamefont {Sikes}, \citenamefont {Jiang}, \citenamefont {Klima}, \citenamefont {Fudenberg}, \citenamefont {Hone}, \citenamefont {Kim},\ and\ \citenamefont {Stormer}}]{Bolotin.2008}%
  \BibitemOpen
  \bibfield  {author} {\bibinfo {author} {\bibfnamefont {K.~I.}\ \bibnamefont {Bolotin}}, \bibinfo {author} {\bibfnamefont {K.~J.}\ \bibnamefont {Sikes}}, \bibinfo {author} {\bibfnamefont {Z.}~\bibnamefont {Jiang}}, \bibinfo {author} {\bibfnamefont {M.}~\bibnamefont {Klima}}, \bibinfo {author} {\bibfnamefont {G.}~\bibnamefont {Fudenberg}}, \bibinfo {author} {\bibfnamefont {J.}~\bibnamefont {Hone}}, \bibinfo {author} {\bibfnamefont {P.}~\bibnamefont {Kim}},\ and\ \bibinfo {author} {\bibfnamefont {H.~L.}\ \bibnamefont {Stormer}},\ }\bibfield  {title} {\bibinfo {title} {Ultrahigh electron mobility in suspended graphene},\ }\href {https://doi.org/10.1016/j.ssc.2008.02.024} {\bibfield  {journal} {\bibinfo  {journal} {Solid State Communications}\ }\textbf {\bibinfo {volume} {146}},\ \bibinfo {pages} {351} (\bibinfo {year} {2008})}\BibitemShut {NoStop}%
\bibitem [{\citenamefont {Golombek}\ \emph {et~al.}(2021)\citenamefont {Golombek}, \citenamefont {Breuer}, \citenamefont {Danzig}, \citenamefont {Kucharczyk}, \citenamefont {Schleberger}, \citenamefont {Sokolowski-Tinten},\ and\ \citenamefont {Wucher}}]{Golombek.2021}%
  \BibitemOpen
  \bibfield  {author} {\bibinfo {author} {\bibfnamefont {A.}~\bibnamefont {Golombek}}, \bibinfo {author} {\bibfnamefont {L.}~\bibnamefont {Breuer}}, \bibinfo {author} {\bibfnamefont {L.}~\bibnamefont {Danzig}}, \bibinfo {author} {\bibfnamefont {P.}~\bibnamefont {Kucharczyk}}, \bibinfo {author} {\bibfnamefont {M.}~\bibnamefont {Schleberger}}, \bibinfo {author} {\bibfnamefont {K.}~\bibnamefont {Sokolowski-Tinten}},\ and\ \bibinfo {author} {\bibfnamefont {A.}~\bibnamefont {Wucher}},\ }\bibfield  {title} {\bibinfo {title} {Generation of ultrashort ke{V} $\mathrm{Ar^{+}}$ ion pulses via femtosecond laser photoionization},\ }\href {https://doi.org/10.1088/1367-2630/abe443} {\bibfield  {journal} {\bibinfo  {journal} {New Journal of Physics}\ }\textbf {\bibinfo {volume} {23}},\ \bibinfo {pages} {033023} (\bibinfo {year} {2021})}\BibitemShut {NoStop}%
\bibitem [{\citenamefont {Kalkhoff}\ \emph {et~al.}(2023)\citenamefont {Kalkhoff}, \citenamefont {Golombek}, \citenamefont {Schleberger}, \citenamefont {Sokolowski-Tinten}, \citenamefont {Wucher},\ and\ \citenamefont {Breuer}}]{Kalkhoff.2023}%
  \BibitemOpen
  \bibfield  {author} {\bibinfo {author} {\bibfnamefont {L.}~\bibnamefont {Kalkhoff}}, \bibinfo {author} {\bibfnamefont {A.}~\bibnamefont {Golombek}}, \bibinfo {author} {\bibfnamefont {M.}~\bibnamefont {Schleberger}}, \bibinfo {author} {\bibfnamefont {K.}~\bibnamefont {Sokolowski-Tinten}}, \bibinfo {author} {\bibfnamefont {A.}~\bibnamefont {Wucher}},\ and\ \bibinfo {author} {\bibfnamefont {L.}~\bibnamefont {Breuer}},\ }\bibfield  {title} {\bibinfo {title} {Path to ion-based pump-probe experiments: Generation of 18 picosecond ke{V} {N}e$^+$ ion pulses from a cooled supersonic gas beam},\ }\href {https://doi.org/10.1103/PhysRevResearch.5.033106} {\bibfield  {journal} {\bibinfo  {journal} {Physical Review Research}\ }\textbf {\bibinfo {volume} {5}},\ \bibinfo {pages} {033106} (\bibinfo {year} {2023})}\BibitemShut {NoStop}%
\bibitem [{\citenamefont {Wiley}\ and\ \citenamefont {McLaren}(1955)}]{Wiley.1955}%
  \BibitemOpen
  \bibfield  {author} {\bibinfo {author} {\bibfnamefont {W.~C.}\ \bibnamefont {Wiley}}\ and\ \bibinfo {author} {\bibfnamefont {I.~H.}\ \bibnamefont {McLaren}},\ }\bibfield  {title} {\bibinfo {title} {Time-of-flight mass spectrometer with improved resolution},\ }\href {https://doi.org/10.1063/1.1715212} {\bibfield  {journal} {\bibinfo  {journal} {Review of Scientific Instruments}\ }\textbf {\bibinfo {volume} {26}},\ \bibinfo {pages} {1150} (\bibinfo {year} {1955})}\BibitemShut {NoStop}%
\bibitem [{\citenamefont {Galanti}\ \emph {et~al.}(1971)\citenamefont {Galanti}, \citenamefont {Gott},\ and\ \citenamefont {Renaud}}]{Galanti.1971}%
  \BibitemOpen
  \bibfield  {author} {\bibinfo {author} {\bibfnamefont {M.}~\bibnamefont {Galanti}}, \bibinfo {author} {\bibfnamefont {R.}~\bibnamefont {Gott}},\ and\ \bibinfo {author} {\bibfnamefont {J.~F.}\ \bibnamefont {Renaud}},\ }\bibfield  {title} {\bibinfo {title} {A high resolution, high sensitivity channel plate image intensifier for use in particle spectrographs},\ }\href {https://doi.org/10.1063/1.1685013} {\bibfield  {journal} {\bibinfo  {journal} {Review of Scientific Instruments}\ }\textbf {\bibinfo {volume} {42}},\ \bibinfo {pages} {1818} (\bibinfo {year} {1971})}\BibitemShut {NoStop}%
\bibitem [{\citenamefont {Blase}\ \emph {et~al.}(2017)\citenamefont {Blase}, \citenamefont {Benke}, \citenamefont {Miller}, \citenamefont {Pickens},\ and\ \citenamefont {Waite}}]{Blase.2017}%
  \BibitemOpen
  \bibfield  {author} {\bibinfo {author} {\bibfnamefont {R.~C.}\ \bibnamefont {Blase}}, \bibinfo {author} {\bibfnamefont {R.~R.}\ \bibnamefont {Benke}}, \bibinfo {author} {\bibfnamefont {G.~P.}\ \bibnamefont {Miller}}, \bibinfo {author} {\bibfnamefont {K.~S.}\ \bibnamefont {Pickens}},\ and\ \bibinfo {author} {\bibfnamefont {J.~H.}\ \bibnamefont {Waite}},\ }\bibfield  {title} {\bibinfo {title} {Microchannel plate detector detection efficiency to monoenergetic electrons between 3 and 28 ke{V}},\ }\href {https://doi.org/10.1063/1.4983338} {\bibfield  {journal} {\bibinfo  {journal} {Review of Scientific Instruments}\ }\textbf {\bibinfo {volume} {88}},\ \bibinfo {pages} {053302} (\bibinfo {year} {2017})}\BibitemShut {NoStop}%
\bibitem [{\citenamefont {Brehm}\ \emph {et~al.}(1995)\citenamefont {Brehm}, \citenamefont {Grosser}, \citenamefont {Ruscheinski},\ and\ \citenamefont {Zimmer}}]{Brehm.1995}%
  \BibitemOpen
  \bibfield  {author} {\bibinfo {author} {\bibfnamefont {B.}~\bibnamefont {Brehm}}, \bibinfo {author} {\bibfnamefont {J.}~\bibnamefont {Grosser}}, \bibinfo {author} {\bibfnamefont {T.}~\bibnamefont {Ruscheinski}},\ and\ \bibinfo {author} {\bibfnamefont {M.}~\bibnamefont {Zimmer}},\ }\bibfield  {title} {\bibinfo {title} {Absolute detection efficiencies of a microchannel plate detector for ions},\ }\href {https://doi.org/10.1088/0957-0233/6/7/015} {\bibfield  {journal} {\bibinfo  {journal} {Measurement Science and Technology}\ }\textbf {\bibinfo {volume} {6}},\ \bibinfo {pages} {953} (\bibinfo {year} {1995})}\BibitemShut {NoStop}%
\bibitem [{\citenamefont {{Y. Yao, A. Metzlaff, A. Wucher, M. Schleberger, L. Breuer, and A. Schleife}}(2023)}]{Y.YaoA.MetzlaffL.BreuerandA.Schleife.2023}%
  \BibitemOpen
  \bibfield  {author} {\bibinfo {author} {\bibnamefont {{Y. Yao, A. Metzlaff, A. Wucher, M. Schleberger, L. Breuer, and A. Schleife}}},\ }\bibfield  {title} {\bibinfo {title} {Real-time electron emission dynamics of graphene under proton irradiation},\ }\href@noop {} {\bibfield  {journal} {\bibinfo  {journal} {arXiv}\ } (\bibinfo {year} {2023})}\BibitemShut {NoStop}%
\bibitem [{\citenamefont {Skopinski}\ \emph {et~al.}(2023)\citenamefont {Skopinski}, \citenamefont {Kretschmer}, \citenamefont {Ernst}, \citenamefont {Herder}, \citenamefont {Madau{\ss}}, \citenamefont {Breuer}, \citenamefont {Krasheninnikov},\ and\ \citenamefont {Schleberger}}]{Skopinski.2023}%
  \BibitemOpen
  \bibfield  {author} {\bibinfo {author} {\bibfnamefont {L.}~\bibnamefont {Skopinski}}, \bibinfo {author} {\bibfnamefont {S.}~\bibnamefont {Kretschmer}}, \bibinfo {author} {\bibfnamefont {P.}~\bibnamefont {Ernst}}, \bibinfo {author} {\bibfnamefont {M.}~\bibnamefont {Herder}}, \bibinfo {author} {\bibfnamefont {L.}~\bibnamefont {Madau{\ss}}}, \bibinfo {author} {\bibfnamefont {L.}~\bibnamefont {Breuer}}, \bibinfo {author} {\bibfnamefont {A.~V.}\ \bibnamefont {Krasheninnikov}},\ and\ \bibinfo {author} {\bibfnamefont {M.}~\bibnamefont {Schleberger}},\ }\bibfield  {title} {\bibinfo {title} {Velocity distributions of particles sputtered from supported two-dimensional {MoS}$_2$ during highly charged ion irradiation},\ }\href {https://doi.org/10.1103/PhysRevB.107.075418} {\bibfield  {journal} {\bibinfo  {journal} {Physical Review B}\ }\textbf {\bibinfo {volume} {107}},\ \bibinfo {pages} {075418} (\bibinfo {year} {2023})}\BibitemShut {NoStop}%
\bibitem [{\citenamefont {Creutzburg}\ \emph {et~al.}(2020)\citenamefont {Creutzburg}, \citenamefont {Schwestka}, \citenamefont {Niggas}, \citenamefont {Inani}, \citenamefont {Tripathi}, \citenamefont {George}, \citenamefont {Heller}, \citenamefont {Kozubek}, \citenamefont {Madau{\ss}}, \citenamefont {McEvoy}, \citenamefont {Facsko}, \citenamefont {Kotakoski}, \citenamefont {Schleberger}, \citenamefont {Turchanin}, \citenamefont {Grande}, \citenamefont {Aumayr},\ and\ \citenamefont {Wilhelm}}]{Creutzburg.2020}%
  \BibitemOpen
  \bibfield  {author} {\bibinfo {author} {\bibfnamefont {S.}~\bibnamefont {Creutzburg}}, \bibinfo {author} {\bibfnamefont {J.}~\bibnamefont {Schwestka}}, \bibinfo {author} {\bibfnamefont {A.}~\bibnamefont {Niggas}}, \bibinfo {author} {\bibfnamefont {H.}~\bibnamefont {Inani}}, \bibinfo {author} {\bibfnamefont {M.}~\bibnamefont {Tripathi}}, \bibinfo {author} {\bibfnamefont {A.}~\bibnamefont {George}}, \bibinfo {author} {\bibfnamefont {R.}~\bibnamefont {Heller}}, \bibinfo {author} {\bibfnamefont {R.}~\bibnamefont {Kozubek}}, \bibinfo {author} {\bibfnamefont {L.}~\bibnamefont {Madau{\ss}}}, \bibinfo {author} {\bibfnamefont {N.}~\bibnamefont {McEvoy}}, \bibinfo {author} {\bibfnamefont {S.}~\bibnamefont {Facsko}}, \bibinfo {author} {\bibfnamefont {J.}~\bibnamefont {Kotakoski}}, \bibinfo {author} {\bibfnamefont {M.}~\bibnamefont {Schleberger}}, \bibinfo {author} {\bibfnamefont {A.}~\bibnamefont {Turchanin}}, \bibinfo {author} {\bibfnamefont {P.~L.}\ \bibnamefont {Grande}}, \bibinfo {author} {\bibfnamefont
  {F.}~\bibnamefont {Aumayr}},\ and\ \bibinfo {author} {\bibfnamefont {R.~A.}\ \bibnamefont {Wilhelm}},\ }\bibfield  {title} {\bibinfo {title} {Vanishing influence of the band gap on the charge exchange of slow highly charged ions in freestanding single-layer {MoS}$_2$},\ }\href {https://doi.org/10.1103/PhysRevB.102.045408} {\bibfield  {journal} {\bibinfo  {journal} {Physical Review B}\ }\textbf {\bibinfo {volume} {102}},\ \bibinfo {pages} {045408} (\bibinfo {year} {2020})}\BibitemShut {NoStop}%
\end{thebibliography}%

\end{document}